\newcommand{\be}{\begin{equation}}
\newcommand{\ee}{\end{equation}}
\newcommand{\bea}{\begin{eqnarray}}
\newcommand{\eea}{\end{eqnarray}}

\newcommand{\br}{{\bf r}}

\newcommand{\bw}{{\bf w}}

\newcommand{\bfs}{{\bf f}}

\newcommand{\bR}{{\bf R}}

\newcommand{\bei}{\begin{itemize}}
\newcommand{\ei}{\end{itemize}}
\newcommand{\bra}{\langle}
\newcommand{\ket}{\rangle}
\newcommand{\dbra}{\langle\!\langle}
\newcommand{\dket}{\rangle\!\rangle}

\newcommand{\bx}{{\bf x}}
\newcommand{\e}{\mbox{\rm e}}
\newcommand{\bsh}{\begin{shadebox}}
\newcommand{\esh}{\end{shadebox}}
\newcommand{\besh}{\mbox{ }\\ \begin{shadebox}}
\newcommand{\eesh}{\end{shadebox}\\ \mbox{ }}
\newcommand{\non}{\nonumber}

\newcommand{\bel}{\begin{list}{$\bullet$}{
\setlength{\itemsep}{0pt}
\setlength{\topsep}{5pt}}}
\newcommand{\el}{\end{list}}
\newcommand{\dg}{\dot{\gamma}}

\newcommand{\bec}{\begin{center}}
\newcommand{\eec}{\end{center}}
\newcommand{\ben}{\begin{enumerate}}
\newcommand{\een}{\end{enumerate}}

\newcommand{\gnear}{
\unitlength=1cm
\begin{picture}(0.56,0)
\put(0.2,0.15){\makebox(0.17,0){$>$}}
\put(0.2,-0.08){\makebox(0.17,0){{\small $\sim$}}}
\end{picture}}

\newcommand{\lnear}{
\unitlength=1cm
\begin{picture}(0.56,0)
\put(0.2,0.15){\makebox(0.17,0){$<$}}
\put(0.2,-0.08){\makebox(0.17,0){{\small $\sim$}}}
\end{picture}}

\makeatletter

\def\eqnarray{
  \stepcounter{equation}
  \let\@currentlabel=\theequation
  \global\@eqnswtrue
  \global\@eqcnt\z@
  \tabskip\@centering
  \let\\=\@eqncr
  $$\halign to \displaywidth\bgroup\@eqnsel\hskip\@centering
  $\displaystyle\tabskip\z@{##}$&\global\@eqcnt\@ne
  \hfill$\displaystyle{{}##{}}$\hfill
  &\global\@eqcnt\tw@$\displaystyle\tabskip\z@{##}$\hfill
  \tabskip\@centering&\llap{##}\tabskip\z@\cr}

\makeatother

\documentclass[aps,prb,twocolumn,twoside,eqsecnum,showpacs,showkeys,floatfix]{revtex4}
\usepackage{graphicx}
\begin{document}

\title{
Analysis of Shear-Thickening in Physical Gel. \\
A Stochastic Theory for Polymer Networks
}

\author{T.~Indei}
\author{T.~Arimitsu}

\affiliation
{Institute of Physics, University of Tsukuba, Ibaraki 305-8571, Japan}

\date{\today}

\begin{abstract}
A formula of steady shear viscosity is derived by introducing a model
to describe the dynamics of physically cross-linked network (physical gel),
and successfully analyzes the shear-thickening
behavior observed in HEUR aqueous 
solutions by Jenkins, Sileibi and El-Aasser. 
We take into account the effects of looped chains at junctions
which detach their one end from the junction as the shear rate increases 
due to the collisions with other chains.
This process produces the weak and wide enhancement of the 
number of chains whose both ends stick to the separate junctions
(active chains) leading to the weak increase in the shear viscosity.
It is also shown that the nonlinear force sustained by 
active chains induces the strong and sharp enhancement of 
the steady shear viscosity.
\end{abstract}



\maketitle


\section{Introduction}  
\label{insection}

One observes generally that the viscosity of polymer 
solution monotonically decreases when 
one increases shear rate.
This effect, referred to as {\it shear-thinning}, results from
the disentanglement of polymer chains under the shear flow.
However, some sort of polymer solutions exhibit an unusual behavior, called 
{\it shear-thickening}, i.e., the viscosity of the solution grows 
as the shear rate increases, attains a maximum and decreases
at higher shear rates.

The shear-thickening phenomenon is observed for 
polymers having a few segments capable of association, such as
hydrophilic polymers containing
a few hydrophobic 
groups,~\cite{Jenkins1,Annable1,Yekta2,Yekta3,Yekta5,Tam,Meins1} 
and ionomers.~\cite{Maus,Cooper}
In solution, these associative polymers construct a
physically cross-linked network due to temporary connections among their 
associative segments (physical gel).
A typical example of such a physical gel
is HEUR (hydrophobically modified ethoxylated urethane)
aqueous solution.~\cite{Jenkins1,Annable1,Yekta2,Yekta3,Yekta5,Tam,Meins1} 
HEUR is a hydrophilic poly(ethylene oxide)
containing hydrophobic 
hydrocarbon groups at its both ends.
In water, the hydrocarbon groups 
at their ends associate 
with each other due to hydrophobic interactions among them to
construct a physically cross-linked network 
at relatively low polymer concentrations.
Another example is
a solution of telechelic ionomers in apolar solvent,~\cite{Maus,Cooper}
such as a solution of $\alpha,\omega$-Mg carboxylato-polyisoprene in 
decahydronaphthalene.~\cite{Maus}

\begin{figure}
\begin{center}
\includegraphics*[bb=35 193 489 539,scale=0.4]{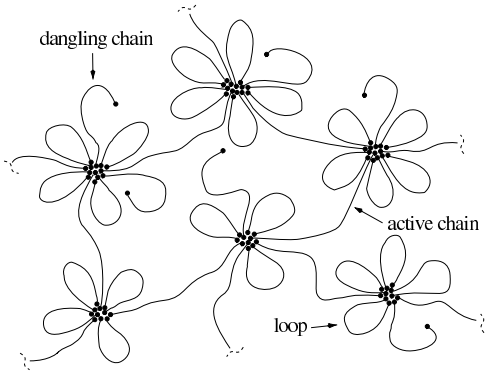}
\end{center}
\caption{
Schematic representation of a network constructed by
HEUR in aqueous solution.\protect\cite{Yekta2}} 
\label{net2}
\end{figure}

In this paper, we study linear chains having 
associative functional groups at both their ends, and analyze the steady 
shear viscosity of HEUR aqueous solutions 
observed by Jenkins {\it et al.}~\cite{Jenkins1}
representing shear-thickening.\cite{phD}
We will consider three types of chains in the model, i.e., 
{\it active chains} with its both ends connecting to separate junctions, 
{\it dangling chains} with only one of its end sticking to a junction
and {\it loops} with both ends connecting to a junction 
(Figure \ref{net2}).
Floating (free) chains in solvent are not taken into account 
in the present approach.
We introduce loops since HEUR in water has been observed to form 
flowerlike micelles comprised primarily of loops,
as schematically depicted in Figure \ref{net2},
for broad polymer concentrations above the onset of association.\cite{Yekta2} 
A dangling chain can become either an active chain 
or a loop by connecting, respectively, 
its one free end to a junction with which
its other end does not stick or has already stuck. In other words,
an active chain can become a loop via dangling state and vice verse.
We take into account the breaking of a loop by detaching its
one end from a junction due to the collisions with
dangling chains or other loops as the shear rate increases.
It will be seen that this process produces the weak and wide 
enhancement of the number of active chains, 
leading to the weak increase in the shear viscosity.
The idea of increase in the number of active chains
around the shear-thickening region
has been suggested from an observation  
that the plateau modulus there
is larger than that in the Newtonian region.\cite{Tam}  
It will be also shown that the nonlinear force 
of active chains described by the inverse Langevin function induces
the strong and sharp enhancement of the steady shear viscosity.
The effect of the nonlinear force has been discussed by others 
in different approaches both theoretically\cite{Vra,Mar0} 
and experimentally\cite{Sere}.
The success in the analysis of the shear viscosity in 
HEUR aqueous solutions by the present theoretical formula reveals that
both the increase in the number of active chains and
the nonlinear force of active chains 
are {\it indispensable} to understand the shear-thickening behavior.

Physical gel is a polymer network in which junctions can break and 
recombine due to thermal fluctuation,
i.e., the junctions have the finite lifetime.
The {\it transient network theory},
formed first by Green and Tobolsky,~\cite{GT}
is a powerful tool
to investigate the rheological property of such gels.
They extended the classical model of rubber elasticity so as to allow for
the junctions which break and reform with constant rate independent 
of shear rate. 
Due to the assumptions of the constant 
breaking rate of junctions
and of the Gaussian chains,
they could not succeed to derive any non-linear rheological behavior. 
Lodge~\cite{Lodge} 
generalized the Green-Tobolsky (GT) model by introducing
the distribution in the lifetime of junctions
(the inverse of breaking rate of junctions)
which does not depend on the shear rate.
Like the GT model, 
obtained shear viscosity did not depend on the shear rate,
i.e., there was neither shear-thinning nor shear-thickening.
On the other hand, Yamamoto~\cite{333-2}
evolved the GT model by
introducing the breakage rate of junctions 
which depends on the end-to-end length of 
a section of chain lying between 
two neighboring junctions (partial chain). 
Since the flow changes the end-to-end length of partial chains,
the breakage rate must depend on the rate of flow. 
By assuming that the junction breaks 
whenever the end-to-end length of the partial chain
exceeds a certain critical value under the flow 
and that the polymer chain is Gaussian, 
he showed that shear-thinning appears. 
This is because the number of partial chains 
decreases as the shear rate grows.

With the intention to analyze the physical gel comprised of linear chains
having functional groups 
at their each end, like HEUR, 
Tanaka and Edwards~\cite{TE0,TE1} developed the transient network model
formed by Yamamoto
by introducing two types of chains in the network, i.e.,
the active chain and the dangling chain.
Similar to the Yamamoto model, they adopted the breakage rate depending on
the end-to-end length of active chains. 
Assuming that the breakage rate is an increasing function of
the end-to-end length of an active chain without a cutoff,
and that the active chain is Gaussian with 
linear force, they obtained the shear-thinning behavior,
but could not the shear-thickening. 

Wang~\cite{Wang} extended the Tanaka-Edwards (TE) model by 
introducing free chains with both ends being not connected to any junctions
in addition to active and dangling chains.
Like Tanaka and Edwards, he adopted the increasing breakage rate of 
an active chain 
without introducing a cutoff to
its end-to-end length. 
He assumed that shear flow promotes creation of 
more free chains which collide with junctions
resulting in the growth of 
the transition rate from a free chain to a dangling chain.
Hence, the number of dangling chains
grows as the shear rate increases, leading to 
the enhancement of the number of active chains
which induces shear-thickening.
However, the adoption of this transition rate appears to be 
somewhat 
doubtful, since
it may not be affected by the frequency of collisions of
a free chain with junctions but affected by
the number of junctions 
surrounding the ends of free chains.\footnote{
The latter effect has been taken into account through the rate equation
$\mu\nu^f\!=\!q\nu^d$~\cite{Wang} within the model, where
$\nu^f$ ($\nu^d$) is the number of free (dangling) chains
in unit volume, and $\mu$ ($q$) is the transition rate from a free (dangling)
to a dangling (free) chain.}

All the transient network models mentioned above have one common assumption,
i.e., the relaxation time of dangling (and free) chains is much
smaller than that of the network.
Under the assumption, a dangling 
chain immediately 
relaxes to, say, the Gaussian chain 
after its connected end detaches from a junction.
Vaccaro and Marrucci~\cite{Mar1}
considered the network model composed of active and dangling chains,
however, in order to avoid the above assumption, they
introduced a `Fokker-Planck equation'
for the distribution of the end-to-end vector of dangling chains.
They adopted the nonlinear 
Warner force\cite{War} for active chains,
and introduced the breakage rate of active chains
which increases with their end-to-end length.
They also introduced the
recombination rate linearly increasing 
with respect to the end-to-end length of dangling chains.
For the relaxation time $\tau$ of chains
having the order of
the lifetime (at equilibrium state) $\beta_0^{-1}$ of active chains, 
shear-thickening is shown to appear.
On the other hand, as seen in subsection \ref{rmatan2}, 
the relaxation time of chains in the experiment 
we are going to analyze\cite{Jenkins1} is estimated as
$\tau\!\simeq\!10^{-5}\!\sim\!10^{-4}$s, 
while $\beta_0^{-1}\!\!\simeq\!10^{-2}\!\sim\!10^{-1}$s. 
For such short 
relaxation time ($\tau\!\ll\!\beta_0^{-1}$), 
shear-thickening does not seem to appear within their model.
Therefore,
we need to propose other mechanism to explain the shear-thickening phenomenon
even in the case $\tau\!\ll\!\beta_0^{-1}$.

The transient network theory may be also applicable to
the topologically entangled polymer system
when the entanglement points are regarded as junctions
since the entanglements formed by chains 
can be disentangled due to the thermal agitation of chains 
along the tube-like region comprised of surrounding chains, i.e., 
reptation proposed by de Gennes.~\cite{dege0}
Actually, Yamamoto evolved his model with the intention to treat 
polymer solutions at rather high concentration.
However, 
the topological restriction of chains is only 
effectively taken into account in this treatment.
Doi and Edwards~\cite{DE0} developed a theory for 
the concentrated polymer solutions or melts on the basis of 
the reptation model.
They succeeded, for example, to derive 
shear-thinning behavior for a topologically entangled system
comprised of linear chains. 

In the present paper, we will focus our attention on the system
where the above effect of the topological entanglement is week and negligible,
since in the experiment we will analyze, the polymer concentration
is at most on the order of the overlap threshold of polymer chains
(see subsection \ref{nphisec}).
Junctions are ascribed to the association among functional groups
at each end of linear chains.

In section \ref{baseeq}, we introduce the basic equation 
dealing with the physically cross-linked network composed of unentangled 
linear chains with associative functional groups at both their ends.
The derived equation, 
representing the transition among each type of 
the chains, will be used throughout in this paper
for the analysis of 
the steady shear viscosity of physical gel.
In section \ref{characchap},
the dynamics of each type of chains
and the transition rates among 
them under the steady shear flow will be discussed. 
We see that the transition from a loop to a dangling chain 
is promoted by the shear flow. 
It plays an important role in the shear-thickening phenomena
as well as the nonlinear force of active chains.
In section \ref{analy}, we solve the 
basic equation introduced in section \ref{baseeq} in order to
give a formula for the steady shear viscosity.
In section \ref{results}, 
we analyze the experimentally observed shear viscosity of 
HEUR aqueous solutions\cite{Jenkins1} by the obtained formula.
It will be shown that the formula explains
experiments quite well for the broad shear rate.
Discussions will be devoted to section \ref{discoz}.

\section{Basic Equation}
\label{baseeq}

\subsection{Assumptions}

We treat in this paper the system made up of linear chains 
which are uniformly distributed in space,
with uniform molecular weight, carrying associative groups at both their ends.
No topological entanglements are considered.
Here we list again the three types of chains relevant to our model
(Figure \ref{net2}):
1) {\it Active chain}: The both ends of the chain 
are connected to two junctions separated in space.
2) {\it Dangling chain}: Only one of the two ends is connected to a junction.
3) {\it Loop}: Both ends are connected to a single junction.

Suppose the situation in which
an incompressible macro-deformation is added to the uniform system.
We assume that the distribution of chains remains uniform
even after the application of 
the deformation. 
Hereafter, we will pay our attention to those phenomena taking place 
in an {\it arbitrary} unit volume in the system.
The number of chains 
$n$ per unite volume is constant in time thanks
to the incompressibility of the deformation, and in space
due to the homogeneity of the system.

The added deformation produces a velocity-field in the system causing
a conformation change of chains.
We represent the conformation of a chain by means of its end-to-end 
vector.


\subsection{End-to-End Vector Space}

Now, we introduce a space, named the {\it end-to-end vector space},
in which we will construct a basic equation for 
the end-to-end vector of chains.
In this space, all chains which have the same end-to-end vector 
are represented by one point called a {\it representative point} 
for the chains. 
Since our system consists of those chains which have the same contour length, 
say $l$, all representative points locate within the sphere of radius $l$. 
Note that the representative point have nothing to do with a position of chain 
in real space. Even if a chain moves around in real space,
its representative point does not move 
unless the chain changes its end-to-end vector.

Thermodynamical quantities related to a chain, such as 
tension, free energy and so on, are often given by functions of 
its end-to-end vector $\br$.
The end-to-end vector space provides us with the support of these functions.


\subsection{Equation for the Number of Chains} 
\label{basic}

\subsubsection{Eulerian Description}
\label{eulabas}

Let us introduce, in unit volume in real space, 
the number $\phi^i(\br,t)d\br$ of $i$-chains 
with end-to-end vector $\br\!\sim\!\br+d\br$ at time $t$.
The superscript $i$ distinguishes the type of chains, i.e.,
$i\!=\!a$ represents active chains, $i\!=\!d$ dangling chains
and $i\!=\!l$ loops.
Note that the total number $\phi(\br,t)d\br$ of chains 
with end-to-end vector $\br\!\sim\!\br+d\br$ 
in unit volume is given by
\bea
\phi(\br,t)d\br=\sum_i\phi^i(\br,t)d\br. \label{warerrerer22} 
\eea
We derive the equation for 
$\phi^i(\br,t)$ in the end-to-end vector space for $i$-chains
under the velocity-field $\dot{\br}^i(\br,t)$
caused by the deformation added to the system.

\begin{figure}[h]
\begin{center}
\includegraphics*[scale=0.9]{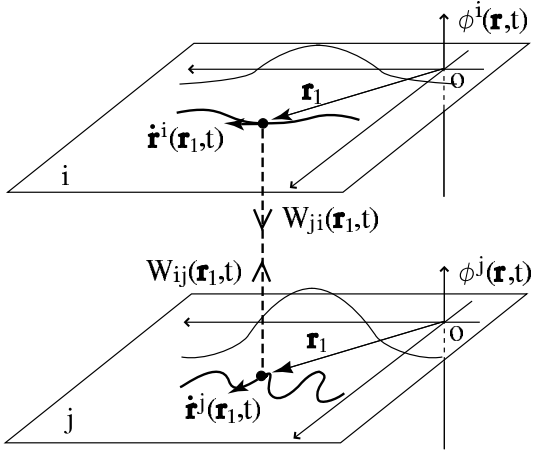}
\end{center}
\caption{
Schematic explanation of transitions
between $i$-chains and $j$-chains
having the same end-to-end vector $\br_1$ at time $t$ 
(Eulerian description).
The upper plane (actually three dimensional)
stands for the end-to-end vector space for $i$-chains
and the lower for $j$-chains.
A solid line in 
the end-to-end vector space for $\alpha$-chains ($\alpha\!=\!i,j$)
represents a streamline for $\alpha$-chains passing through 
a point $\br_1$ at $t$.
The streamline gives the velocity $\dot{\br}^{\alpha}(\br,t)$ 
as its tangent line at $\br$.
A representative point for $\alpha$-chains moves in 
accordance with $\dot{\br}^{\alpha}(\br, t)$ 
in each end-to-end vector space as described by (\protect\ref{ware33}).}
\label{ijpicteu}
\end{figure}

Let us consider a volume $V^i(t)$,
in the end-to-end vector space for $i$-chains,
which varies its boundary in accordance with $\dot{\br}^i(\br,t)$.
Since the points for $i$-chains in this volume
are created or annihilated, 
we have the equation for number-conservation of $i$-chains
\bea
\frac{d}{dt}\Biggl( \int_{V^i(t)} \!\!\!\!d\br~ \phi^i(\br,t) \Biggr) 
&=&-\sum_{j(\neq i)}\int_{V^i(t)}\!\!\!\!d\br~W_{ji}(\br,t)\phi^i(\br,t) \non\\
& & \hspace*{-0.8cm}
+\sum_{j(\neq i)}\int_{V^i(t)} \!\!\!\!d\br~ W_{ij}(\br,t)\phi^j(\br,t),
\label{warenama3} 
\eea
where $W_{ij}(\br,t)dt$ is the transition probability during $dt$
at time $t$ from $j$-chains to $i$-chains
having the same end-to-end vector $\br$ (see Figure \ref{ijpicteu}).
The conservation of the transition probability $W_{ij}(\br,t)dt$ reads
\bea
\sum_{i}W_{ij}(\br,t)dt=1.
\eea
The first term in the right-hand side of (\ref{warenama3}) 
represents the number of $i$-chains 
in $V^i(t)$ which convert to other types of chains in unit time,
whereas the second term stands for the number of 
$j(\neq\! i)$-chains
in $V^i(t)$ which convert to $i$-chains in unit time.
That is, the first and the second terms represent annihilation and 
creation of $i$-chains, respectively.
As $V^i(t)$ is arbitrary, 
we obtain 
the equation for $\phi^i(\br,t)$ from (\ref{warenama3})
\bea
& &\frac{\partial}{\partial t} \phi^i(\br,t)
+\frac{\partial}{\partial \br}\cdot\Biggl( \dot{\br}^i
(\br,t)\phi^i(\br,t) \Biggr) \non\\
& &= -\sum_{j(\neq i)}W_{ji}(\br)\phi^i(\br,t)
+\sum_{j(\neq i)} W_{ij}(\br)\phi^j(\br,t).\label{ware33}
\eea

When the {\it total} number of chains does not change,
$\phi(\br,t)$ satisfies an equation
\bea 
\frac{\partial}{\partial t}\phi(\br,t)
+\frac{\partial}{\partial\br}\cdot
\Biggl( \dot{\br}(\br,t)\phi(\br,t) \Biggr)=0. \label{ststs}
\eea
The requirement that the summation of (\ref{ware33})
with respect to $i$ should 
coincide with (\ref{ststs}) gives us the relation between
$\dot{\br}(\br,t)$ and $\dot{\br}^i(\br,t)$ in the form
\bea
\dot{\br}(\br,t)\phi(\br,t)=\sum_i
\dot{\br}^i(\br,t)\phi^i(\br,t). \label{rdwaaa33}
\eea

\subsubsection{Lagrangian Description}
\label{lagagebas}

The basic equation (\ref{ware33}) of the present paper
is the one in the Eulerian description.
Since, hereafter, we will analyze mainly in the Lagrangian description,
we introduce the quantity
\bea
\hspace*{-0.4cm}
\phi^i(\br^i(t;\br_0^i,t_0),t)\!=\!\!\!\int\!\! 
d\br~ \delta(\br-\br^i(t;\br_0^i,t_0))
\phi^i(\br,t),\label{kd2}
\eea
in the description
where $\br^i(t;\br_0^i,t_0)$ is a position of the representative point
belonging to $i$-chains at $t$, 
which was located at $\br_0^i$ at the initial time $t_0$.
$\br^i(t;\br_0^i,t_0)$ is a solution of the equation of motion
\bea
\frac{d\br^i(t)}{dt}=\dot{\br}^i(\br^i(t),t),\label{reqmot}
\eea
with the initial condition $\br^i(t_0;\br_0^i,t_0)\!=\!\br_0^i$.
The equation of motion (\ref{reqmot})
describes the dynamics of the end-to-end vector in real space,
which corresponds to the movement of a representative point 
under the influence of the flow $\dot{\br}^i(\br,t)$ 
in the end-to-end vector space.

\begin{figure}[ht]
\begin{center}
\includegraphics*[scale=0.9]{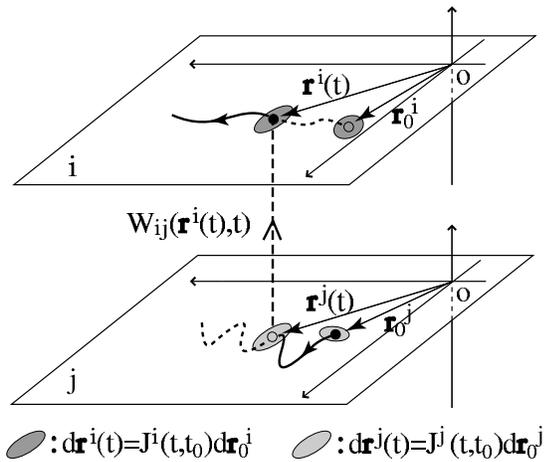} 
\end{center}
\caption{
Schematic explanation of the transition
from $j$-chains to $i$-chains (Lagrangian description).
A solid line in the end-to-end vector space for $i$-chains (upper plane) 
expresses a trajectory of a representative point for $i$-chains
which was at $\br_0^i$ at time $t_0$.
On the other hand, 
a solid line in the end-to-end vector space for $j$-chains (lower plane)
represents a trajectory of a representative point for $j$-chains
passing through a point $\br^j(t)$ which is equal to $\br^i(t)$
at the moment $t$.
The dark-gray region at 
$\br^i(t)$ in the upper plane 
represents a domain 
$d\br^i(t)\!=\!J^i(t,t_0)d\br^i_0$
whereas the light-gray region 
around     
$\br^j(t)(=\!\br^i(t))$ 
in the lower plain 
is $d\br^j(t)\!=\!J^j(t,t_0)d\br^j_0$ which is equal to $d\br^i(t)$.}
\label{ijpict}
\end{figure}

Now, we derive the equation for $\phi^i(\br^i(t;\br_0^i,t_0),t)$.
Differentiating (\ref{kd2}) with respect to $t$, we have 
\bea
\lefteqn{\frac{d\phi^i(\br^i(t),t)}{dt}+
\frac{1}{J^i(t,t_0)}\frac{dJ^i(t,t_0)}{dt}
\phi^i(\br^i(t),t)} \non\\
&=&\int \hspace*{-0.1cm}d\br ~\delta(\br-\br^i(t))
\Biggl(\frac{\partial\phi^i(\br,t)}{\partial t}
\!+\!\frac{\partial}{\partial\br}
\!\cdot\!\Biggl( \dot{\br}^i(\br,t)
\phi^i(\br,t)\Biggr) \Biggr),\non\\ \label{lagnotoy}
\eea
where $J^i(t,t_0)$ is the Jacobian defined by
\bea
J^i(t,t_0)=\Biggl|\frac{\partial r_{\alpha}^i(t)}
{\partial r_{0\hspace*{0.2mm}\beta}^i}\Biggr|.
\eea
For simplicity, we are omitting $\br_0^i$ and $t_0$ 
in $\br^i(t;\br_0^i,t_0)$. 
By putting the equation (\ref{ware33}) within 
the Eulerian description into (\ref{lagnotoy}) 
and by multiplying $J^i(t,t_0)$ on both sides, (\ref{lagnotoy}) 
reduces to 
\bea
\lefteqn{\frac{d}{dt}
\Biggl( J^i(t,t_0)\phi^i(\br^i(t),t) \Biggr)}\non\\
&=&-\sum_{j(\neq i)}W_{ji}(\br^i(t),t)J^i(t,t_0)\phi^i
(\br^i(t),t)\non\\
& &+\sum_{j(\neq i)} W_{ij}(\br^i(t),t)J^i(t,t_0)
\phi^j(\br^i(t),t).\label{lagnobibun}
\eea
A transition  
to the $i$-chain with 
end-to-end vector $\br^i(t)$ (or to the region 
$d\br^i(t)\!=\!J^i(t,t_0)d\br_0^i$
in the end-to-end vector for $i$-chains)
can occur only from  
the $j$-chains ($j\! \neq \!i$) 
whose end-to-end vector $\br^j(t)$ is equal to $\br^i(t)$ 
(or from the  
end-to-end vector space for $j$-chains
whose region $d\br^j(t)\!=\!J^j(t,t_0)d\br_0^j$ is equal to $d\br^i(t)$).
Therefore, the arguments of $W_{ij}$ and $\phi^j$ 
in the second term of the right-hand side of (\ref{lagnobibun}) 
are $\br^i(t)$ instead of $\br^j(t)$ (or the superscript of the 
Jacobian $J^i(t,t_0)$ is $i$ instead of $j$) (see Figure \ref{ijpict}).

Integrating (\ref{lagnobibun}) from time $t_0$ to $t$,
we have 
\bea
& &J^i(t,t_0)\phi^i(\br^i(t;\br_0^i,t_0),t) \non\\
& &\hspace*{0cm}=\exp\Biggl[ -\int_{t_0}^tdt^{\prime}
\sum_{j(\neq i)}W_{ji}(\br^i(t^{\prime};\br_0^i,t_0),t^{\prime})
\Biggr]\phi^i(\br_0^i,t_0) \non\\
& & \hspace*{0.1cm}+\int_{t_0}^tdt^{\prime}
\exp\Biggl[ -\int_{t^{\prime}}^tdt^{\prime\prime}\sum_{j(\neq i)}
W_{ji}(\br^i(t^{\prime\prime};\br_0^i,t_0),t^{\prime\prime})\Biggr]\non\\
& &\hspace*{0.1cm}
\times \!\sum_{j(\neq i)}W_{ij}(\br^i(t^{\prime};\br_0^i,t_0),t^{\prime})
J^i(t^{\prime},t_0)\phi^j(\br^i(t^{\prime};\br_0^i,t_0),t^{\prime}).
\label{lagsekibun}\non\\
\eea
The first term in the right-hand side of (\ref{lagsekibun})
expresses the number of $i$-chains with end-to-end vector 
$\br^i(t)\!\sim\!\br^i(t)\!+\!d\br^i(t)$ which
were $i$-chain having the end-to-end vector 
$\br_0^i\!\sim\!\br_0^i\!+\!d\br_0^i$ 
at time $t_0$, 
and remain $i$-chain until time $t$. 
This term vanishes at $t\!\to\!\infty$ in most cases.
The second term represents the number of $i$-chains with 
$\br^i(t)\!\sim\!\br^i(t)\!+\!d\br^i(t)$ which became $i$-chain 
at $t^{\prime}$ ($>t_0$), and remain $i$-chain until time $t$.

\subsection{Stochastic Process}
\label{stprosec}

When the dynamics of $i$-chains is given by
a stochastic process, say $\bR(t)$,  
$\dot{\br}^i(\br,t)$ becomes stochastic $\dot{\br}^i(\br,t;\bR(t))$. 
In this case,
$\phi^i(\br,t)$ 
becomes also stochastic $\phi^i(\br,t;\bR(t))$ through (\ref{ware33}).
Hereafter, we denote $\dot{\br}^i(\br,t;\bR(t))$ as $\dot{\br}_f^i(\br,t)$
and $\phi^i(\br,t;\bR(t))$ as $\phi_f^i(\br,t)$ for simplicity.
The equations for $\phi^i(\br,t)$ in the previous subsection 
hold for $\phi_f^i(\br,t)$ as well, 
if the products between two stochastic quantities, such as 
$\dot{\br}_f^i(\br,t)\phi^i_f(\br,t)$,
are regarded as the Stratonovich-type products.~\cite{strat}
By replacing $\phi^i(\br,t)$ and $\dot{\br}^i(\br,t)$ 
with $\phi_f^i(\br,t)$ and $\dot{\br}^i_f(\br,t)$ respectively,
(\ref{ware33}) becomes
\bea
& &\frac{\partial}{\partial t} \phi_f^i(\br,t)
+\frac{\partial}{\partial \br}\cdot\Biggl( \dot{\br}^i_f
(\br,t)\phi_f^i(\br,t) \Biggr) \non\\
& &\hspace*{0.5cm}= -\sum_{j(\neq i)}W_{ji}(\br,t)\phi_f^i(\br,t)
+\sum_{j(\neq i)}W_{ij}(\br,t)\phi_f^j(\br,t),\non\\
\label{ware33ff}
\eea
and by replacing $\phi^i(\br^i(t),t)$ and $\br^i(t)$
with $\phi^i_f(\br^i_f(t),t)$ and $\br^i_f(t)$ respectively,
(\ref{lagsekibun}) becomes
\bea
& &J^i(t,t_0)\phi_f^i(\br^i_f(t;\br_0^i,t_0),t)  \non\\
& &=\exp\Biggl[ -\int_{t_0}^tdt^{\prime}
\sum_{j(\neq i)}W_{ji}(\br^i_f(t^{\prime};\br_0^i,t_0),t^{\prime})
\Biggr]\phi_f^i(\br_0^i,t_0) \non\\
& & +\int_{t_0}^tdt^{\prime}
\exp\Biggl[ -\int_{t^{\prime}}^tdt^{\prime\prime}\sum_{j(\neq i)}
W_{ji}(\br^i_f(t^{\prime\prime};\br_0^i,t_0),t^{\prime\prime})\Biggr]\non\\
& & \times\!\sum_{j(\neq i)}
W_{ij}(\br^i_f(t^{\prime};\br_0^i,t_0),t^{\prime})
J^i(t^{\prime},t_0)\phi_f^j(\br^i_f(t^{\prime};\br_0^i,t_0),t^{\prime}),
\non\\
\label{lagsekibunff}
\eea
where $\br^i_f(t;\br_0^i,t_0)$ is
a solution of the Langevin equation
\bea
\frac{d\br^i(t)}{dt}=\dot{\br}^i(\br^i(t),t;\bR(t))
=\dot{\br}_f^i(\br^i(t),t),
\label{reqmotf}
\eea
with the initial condition $\br^i_f(t_0;\br_0^i,t_0)\!=\!\br_0^i$.
An equation (\ref{ware33ff}) corresponds to the 
`stochastic Liouville equation' in the end-to-end vector space 
for $i$-chains.~\cite{kubo}
In the following, we will denote by $\bra\!\bra\cdots\ket\!\ket$
the average with respect to the stochastic process.

\subsection{Remarks}

In the present model,
features of each type of 
chains are characterized through 
$\dot{\br}^i_f(\br,t)$ and $W_{ij}(\br,t)$. 
We will 
give functional forms of $\dot{\br}^i_f(\br,t)$ and $W_{ij}(\br,t)$
in the next section when the system is in the incompressible 
velocity gradient $\hat{\kappa}(t)$, especially, in the steady shear flow.


\section{Characterization} 
\label{characchap}

\subsection{Tension in a Chain}
\label{tenchara}

When the distance between two ends of a flexible chain is finite, say $\br$,
there appears an effective attractive force $\bfs(\br)$ between these ends, 
due to the thermal motion of segments within the chain.
We adopt, for the attractive force, 
the one obtained within the random-flight model:
\bea
{\bf f}(\br)=-\frac{k_BT}{a}L^{-1}(\frac{r}{Na})~\frac{\br}{r},
\label{langtyoury}
\eea
where $L^{-1}$ is the inverse function of the Langevin function
defined by
$L(x)\!=\!\coth x\!-\!1/x$.
Here and hereafter, $N$ is the number of 
statistical segments
in a chain, $a$
the length of a segment,
$T$ temperature of the solution in which chains are immersed
and $k_B$ the Boltzmann constant.
Note that in the random-flight model, the finiteness of the 
contour length $l\!=\!Na$ of a 
chain is taken into account seriously.


\subsection{Dynamics in End-to-End Vector Space}
\label{dyna}

\subsubsection{Active Chain}
\label{charac}

We assume that all junctions 
move in accordance with the velocity gradient $\hat{\kappa}(t)$,
and that the effects of the fluctuation of junctions are week and negligible
(affine deformation assumption).
Since the end points $\bx_1^a(t)$ and $\bx_2^a(t)$ of an active chain
in real space are sticking to {\it different} junctions by definition,
the end-to-end vector $\br^a(t)$, defined by 
$\br^a(t)\!=\!\bx^a_1(t)-\bx^a_2(t)$, has finite value whose
time evolution is given by
\bea
\frac{d\br^a(t)}{dt}=\hat{\kappa}(t)\br^a(t). \label{activevnama}
\eea
The equation (\ref{activevnama}) is equivalent to the velocity-field  
\bea
\dot{\br}^{a}(\br,t)=\hat{\kappa}(t)\br \label{activevnaba}
\eea
in the end-to-end vector space for active chains.
It means that
if a representative point at $\br$ belongs to active chains at $t$, 
it flows in accordance with $\dot{\br}^a(\br, t)$.
Integrating (\ref{activevnama}) from time $t_0$ to $t$, we have
\bea
\br^a(t;\br_0^a,t_0)\!
&=&\!\mbox{T}\exp\Biggl[ \int_{t_0}^t \!dt^{\prime}
\hat{\kappa}(t^{\prime}) \Biggr] \br_0^a 
\!\equiv\!\hat{\lambda}(t,t_0)\br_0^a, \label{tnoto3}
\eea
where T is the time-ordered operator. 
In the case of steady flows, since $\hat{\kappa}$ is independent of time,
$\hat{\lambda}(t;t^{\prime})$ becomes a function of the time interval 
$t-t^{\prime}$, i.e., 
$\hat{\lambda}(t;t^{\prime})\!=\!\exp[(t\!-\!t^{\prime})\hat{\kappa}]  
\equiv \hat{\lambda}(t\!-\!t^{\prime})$.

\subsubsection{Dangling Chain}
\label{chardang}

One end of a dangling chain connects to a 
junction (we call it connected end), whereas the other end
does not (we call it free end).
We treat the free end as a Brownian particle, i.e.,
the free end changes its position randomly
by the random force $\bR(t)$ acting on it.
Let $\bx_1^d$ stands for the position of the free end of a dangling chain,
and $\bx_2^d$ the position of its connected end in real space.
Dynamics of $\bx_1^d$ is determined by the Langevin equation
\bea
\hspace*{-0.7cm}
\zeta_N \Biggl(\!
\frac{d\bx^d_1(t)}{dt}-\hat{\kappa}(t)\bx^d_1(t)\!\Biggr) 
=\bfs(\bx^d_1(t)-\bx^d_2(t))+\bR(t),\label{langevinmono} 
\eea
where we neglected the inertial term ($\propto\!d^2\bx^d_1(t)/dt^2$)
since the relaxation time for the velocity of 
typical polymer chains are usually 
much smaller than the time region of our interest.~\cite{doiyaku}
$\zeta_N$ is the friction constant between a chain
and a solvent, which is assumed to be
$\zeta_N\!\simeq N\!\zeta_1$
where $\zeta_1$ is the friction constant per 
segment. $\zeta_1$ is given by Stokes's law: $\zeta_1\!=\!6\pi\eta_s a$ 
with $\eta_s$ being the viscosity of the solvent.
The random force $\bR(t)$ in (\ref{langevinmono})
is assumed to be the Gaussian white process,
i.e., its average and variance are given by
\bea
& &\bra\!\bra R_{\alpha}(t)\ket\!\ket=0 ,\\
& &\bra\!\bra R_{\alpha}(t)R_{\beta}(s)\ket\!\ket=
2\zeta_N k_BT\delta_{\alpha \beta}\delta(t-s),
\eea
respectively $(\alpha,\beta\!=\!x,y,z)$.
The dynamics of $\bx_2^d$ is determined 
in accordance with $\hat{\kappa}(t)$:
\bea
\frac{d\bx^d_2(t)}{dt}=\hat{\kappa}(t)\bx^d_2(t).
\label{ljakot}
\eea
Making use of (\ref{langevinmono}) and (\ref{ljakot}), we obtain the
equation for the time evolution of the end-to-end vector 
$\br^d(t)=\bx^d_1(t)-\bx^d_2(t)$ in the form
\bea
\frac{d\br^d(t)}{dt}=\hat{\kappa}(t)\br^d(t)
+\frac{1}{\zeta_N}\Biggl(\bfs(\br^d(t))+\bR(t)\Biggr).
\label{dkoremono}
\eea
The equation (\ref{dkoremono}) is equivalent to the velocity-field 
\bea
\dot{\br}_f^d(\br,t)=\hat{\kappa}(t)\br
+\frac{1}{\zeta_N}\Biggl(\bfs(\br)+\bR(t)\Biggr)
\label{dkore0}
\eea
in the end-to-end vector space for dangling chains.
It means that
if a representative point at $\br$ belongs to dangling chains at $t$, 
it flows in accordance with $\dot{\br}_f^d(\br, t)$.

When the Hookean approximation is adopted for the tension
$\bfs(\br)$, i.e.,
\bea
{\bf f}(\br)=-K_N\br ~~(\equiv \bfs_N^0(\br)), ~~~~
K_N=\frac{3k_BT}{Na^2}, \label{letia}
\eea 
the equation of motion (\ref{dkoremono}) 
and the corresponding velocity-field (\ref{dkore0}) become
\bea
\frac{d\br^d(t)}{dt}=\Biggl(\hat{\kappa}(t)
-\frac{K_N}{\zeta_N}\Biggr)\br^d(t)+\frac{1}{\zeta_N}\bR(t)
\label{dkoremonew}
\eea
and 
\bea
\dot{\br}_f^d(\br,t)=\Biggl(\hat{\kappa}(t)
-\frac{K_N}{\zeta_N}\Biggr)\br+\frac{1}{\zeta_N}\bR(t),
\label{dkore1}
\eea
respectively. The equation of motion 
(\ref{dkoremonew}) is analytically integrated from $t_0$ to $t$ as
\bea
\br^d_f(t;\br_0^d,t_0)&=&\e^{-(t-t_0)/2\tau}
\hat{\lambda}(t,t_0)\br_0^d \non\\
& &\hspace*{-1cm}+\frac{1}{\zeta_N}\!\int_{t_0}^{t}\!
dt^{\prime}\e^{-(t-t^{\prime})/2\tau}
\hat{\lambda}(t,t^{\prime})\bR(t^{\prime}), \label{dantoketa2}
\eea
where $\tau$ is the relaxation time of a dangling chain defined by
\bea
\tau=\frac{\zeta_N}{2K_N}\simeq\frac{\zeta_1 N^2a^2}{6k_BT}.\label{deftau}
\eea
The Hookean approximation (\ref{letia})
is valid when the end-to-end length of a dangling chain is 
short enough compared with its contour length.
This situation is realized when
the relaxation time (\ref{deftau}) 
for the end-to-end vector of a dangling chain is 
much shorter than the characteristic time of the deformation.

\subsubsection{Loop}
\label{charloop}

The end-to-end distance of a loop is {\it always} zero, i.e.,
\bea
\br^l(t;\br_0^l\!=\!0,t_0)=0, ~~~~~\mbox{(for any $t(>t_0)$).}\label{loopdmo2}
\eea
The equation (\ref{loopdmo2}) 
is equivalent to the velocity-field 
\bea
\dot{\br}^l(\br=0,t)=0,~~~~~\mbox{(for any $t$).} \label{lkore0}
\eea
in the end-to-end vector space for loops.
It means that if a representative point at $\br\!=\!0$ 
belongs to loops at $t$, it does not flow.
Note that no information about the conformation of a chain
is obtained from (\ref{loopdmo2}) or (\ref{lkore0}).

Now, we estimate the dimension of a loop in a steady shear flow.
Let $\bx(n,t)$ be a position of the $n$th segment ($n\!=\!1,\cdots,N$) 
of a chain in real space at time $t$.
The Langevin equation for segments is written, 
for $n\!=\!2,\cdots,N\!-\!1$, as
\bea
& &\zeta_1\Biggl( \frac{d\bx(n,t)}{dt}-
\hat{\kappa}(t)\bx(n,t)\Biggr)\non\\
& &=\bfs_1^0(\bx(n,t)-\bx(n+1,t))+\bfs_1^0(\bx(n,t)-\bx(n-1,t)) \non\\
& &~~+\bR(n,t) \label{lanlosao} \\
& &= K_1\frac{\partial^2 \bx(n,t)}{\partial n^2}+\bR(n,t)
\label{lanlosaren}
\eea
where, in (\ref{lanlosao}), we substituted (\ref{letia})
for the force $\bfs_{N=1}^0$ between neighboring two segments along the chain,
and took the continuous limit for $n$.
The random force $\bR(n,t)$, acting on the  
segment at $\bx(n,t)$,
is assumed to be the Gaussian white process, i.e.,
its average and variance are given by
\bea
& &\hspace*{-0.9cm}\bra\!\bra R_{\alpha}(n,t)\ket\!\ket=0 ,\label{rloopj1}\\
& &\hspace*{-0.9cm}\bra\!\bra R_{\alpha}(n,t)R_{\beta}(m,s)\ket\!\ket\!=\!
2\zeta_1 k_BT\delta_{\alpha \beta}\delta(n-m)\delta(t-s),\label{rloopj2}
\eea
respectively.
Since one end of a loop connects with its another end,
the boundary condition at $n\!=\!0$ and $N\!=\!0$ should be
\bea
\bx(n\!=\!0,t)=\bx(n\!=\!N,t)=0, \label{bankon0}
\eea
where we fixed the connected point at the origin without loss of generality.
The property of a loop is characterized through (\ref{bankon0}).

Making use of (\ref{lanlosaren}) with (\ref{bankon0}),
we can get the {\it mean-square radius of gyration} 
$s_{\alpha}s_{\beta}$ of a loop defined by
\bea
& &\hspace*{-0cm}s_{\alpha}s_{\beta} \non\\
& &\hspace*{-0cm}\!=\frac{1}{N}\!\int_0^N \hspace*{-0.2cm} dn 
\dbra [x_{\alpha}(n,t)-x_{G\alpha}(n,t)]
[x_{\beta}(n,t)-x_{G\beta}(n,t)]\dket  \non\\
\label{loopnpse}
\eea
under a deformation.
Here, $\bx_G(t)\!=\!N^{-1}\!\!\int_0^N dn \bx(n,t)$ is the center 
of mass of a loop.
When a steady shear deformation with shear rate 
$\dg$, described by the deformation tensor
\bea
\hat{\lambda}(t,t^{\prime}) 
=\left( \begin{array}{ccc}
1 & \dot{\gamma}(t-t^{\prime}) & 0 \\
0 & 1 & 0 \\
0 & 0 & 1
\end{array}\right) \label{shearlam}
\eea
or the velocity gradient tensor 
\bea
\hat{\kappa} 
=\left( \begin{array}{ccc}
0 & \dot{\gamma}& 0 \\
0 & 0 & 0 \\
0 & 0 & 0
\end{array}\right) \label{shearlamkap}
\eea
is added, (\ref{loopnpse}) become (see appendix \ref{appenradi})
\bea
& &s_x^2=\frac{Na^2}{36}\Biggl(1+\frac{13\pi^4}{10080}(\dg\tau_1)^2\Biggr),
\label{loopnopx}\\
& &s_y^2=s_z^2=\frac{Na^2}{36}, \label{ynorads}\\
& &s_xs_y=\frac{Na^2}{72}, ~~s_xs_z=s_ys_z=0. \label{ynoradsonota} 
\eea
$\tau_1$ is the relaxation time of a loop given by
\bea
\tau_1 
=\frac{\zeta_1N^2a^2}{3\pi^2k_BT} \label{raustau},
\eea
which  is essentially 
equivalent to $\tau$ defined by (\ref{deftau}).

\subsection{Transition Probabilities}
\label{trasec}

The $j$-chain becomes the $i$-chain
with the probability $W_{ij}$ during unit time.
We suppose that the transition rates $W_{ij}$ are affected by
how many times the $j$-chain collides with other chains 
within its lifetime under the influence of a flow.
Since, in most cases, the frequency of the collision 
depends on the velocity gradient $\hat{\kappa}(t)$ of a flow, 
$W_{ij}$ is a function of $t$ through $\hat{\kappa}(t)$ 
as well as the end-to-end vector $\br$ of the $j$-chain.
However, at steady states in steady flows, $W_{ij}$ becomes a 
function of a quantity characterizing the flow, e.g., 
the shear rate $\dg$ for the steady shear flow.
In this subsection, for the shear flow given by
(\ref{shearlamkap}), we introduce the dependence of $W_{ij}$ on 
$\dg$ and $\br$.

In the following, for simplicity, we denote 
the transition $W_{da}$ from an active to a dangling chain by $\beta$, 
$W_{ad}$ from a dangling to an active chain by $p$, 
$W_{ld}$ from a dangling chain to a loop by $v$ and
$W_{dl}$ from a loop to a dangling chain by $u$.
We can put 
$W_{la}\!=\!W_{al}\!=\!0$ 
between an active chain and a loop, since the primitive transitions
should occur between the dangling and other states
as stated in section \ref{insection}.

\subsubsection{Transition between Active and Dangling Chains}
\label{trabetap}

\paragraph{From Active to Dangling Chains} 

An active chain becomes a dangling chain with the transition rate
$\beta$ by detaching its one end from a junction.
In our model, $\beta$ depends scarcely on
$\dg$. 
The reason is as follows.
{\it All} active chains alter their end-to-end vector {\it affinely}, 
therefore, they hardly collide with each other.
As for the dangling chain (or the loop), since it 
sticks to a certain junction, its center of mass moves
in accordance with $\hat{\kappa}$.
That is, on a certain $xz$-plane (i.e., 
the plane on which the shear flow has the same velocity), 
the relative velocity 
between the active chain and the dangling chain (or the loop) is 
very slow.
This is the reason why the frequency of the collision
between the active chain and the dangling chain (or the loop) 
is small.

We assume that 
the one end of an active chain dissociates 
from the network definitely when its end-to-end distance becomes longer than 
$r^*(\!<\!Na)$.
We further assume that the breakage rate $\beta(\br)$ for $r\le r^*$
is constant and has value $\beta_0$.
The breakage rate $\beta(\br)$ under these assumptions can be written in 
the form\cite{333-2}
\bea	
\beta (\br) = \left\{
           \begin{array}{@{\,}ll}
            \beta_0 & \mbox{($r\le r^*$)} \\
            \infty & \mbox{($r>r^*$)}
            \end{array}
          \right. .\label{betaour0} 
\eea

\paragraph{From Dangling to Active Chains} 

A dangling chain becomes an active chain with the transition rate $p$
by attaching its free end 
to a certain junction with which 
its other end of the chain 
does not connect.
Although the dangling chain collides with other 
dangling chains and loops, there seems to be no reason for
the collision to affect the {\it connecting} process of the 
free end to the junction.
Therefore, we assume that $p$ is independent of $\dg$.
As for the $\br$ dependence,  
we can regard $p$ as a constant 
effectively, since 
the connecting process does not seem to be affected strongly by 
the end-to-end vector.

\subsubsection{Transition between Dangling Chains and Loops} 
\label{loopdangtra}

\paragraph{From Dangling Chains to Loops} 

A dangling chain becomes a loop 
with the transition rate $v$ by attaching its free end
to the junction with which its other end has 
already stuck.
Since the connecting process is not influenced by the collision with 
other chains as stated above, $v(\vec r)$ can be assumed to be independent of 
$\dot{\gamma}$.

Since 
the process takes place only when
its end-to-end length is $0$, we can write $v(\br)$ as
\bea
v(\br)=v_0\delta(\br)V,
\eea
where $V$ is the quantity having the dimension of the volume 
(see (\ref{defVV})).

\paragraph{From Loops to Dangling Chains} 

A loop becomes a dangling chain 
with the transition rate $u$ by detaching its one end
from the junction with which its both ends has 
stuck.
The {\it disconnecting} process of the loop are influenced by
collisions with dangling chains and other loops.

Since the transitions from the {\it active} chain to the dangling chain
is also 
the disconnecting process of one end from a junction,
the breakage rate $u(\dot{\gamma})$
for the loop when $\dg\!=\!0$ should be
the same as the detaching rate for 
the {\it active} chain appeared in (\ref{betaour0}),
i.e., $u(\dg\!=\!0)=\beta_0$.
Let us introduce the quantity 
$u(\dg)/\beta_0$ $(\dg\!\neq\!0)$ 
which stands for the probability
for the loop to dissociate during the time duration $1/\beta_0$.
Note that $1/\beta_0$ stands for the lifetime of loops when $\dg\!=\!0$.
The ratio $u(\dg)/\beta_0$ must be larger than $1$, 
since the collisions of other chains with the backbone of the loop
affects the possibility 
for its one of the ends to disconnect from the junction.
We can assume that the increment $A$ of $u(\dg)/\beta_0$, defined by
\bea
A=\frac{u(\dg)}{\beta_0}-1, 
\label{konkai}
\eea
is given by the number of {\it segments} 
colliding with the loop {\it during its lifetime}.
The collisions take place when the segments enter into
the region occupied by the loop.
We assume that the radius of the region is given by its 
radius of gyration.

\begin{figure}[t]
\begin{center}
\includegraphics*[scale=0.5]{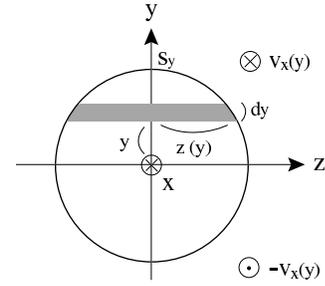}
\end{center}
\caption{Schematic picture of a region occupied by a loop, represented 
as a sphere with a radius of $s_y$.
Monomers in the region $y\!>\!0$ enter 
into this domain along the $x$-axis, 
whereas segments in $y\!<\!0$ along the $-x$ direction.}
\label{nyuusya1fig}
\end{figure}

Under the above assumption, $A$ is estimated as follows.
Let the center of mass of a loop be the origin.
Denoting by $s_y$ the radius of gyration of the loop 
of $y$-component (see (\ref{ynorads})),
the number of segments $\nu_{in}^{d,l}(y)dy$ belonging to 
dangling chains and loops, which enter into a region 
$y\!\sim\!y+dy~(y>0)$ and $-z(y)\!\sim\!z(y)$ occupied by the loop 
(see Figure \ref{nyuusya1fig}) 
in unit time, is given by
\bea
\nu_{in}^{d,l}(y)dy&=&(\nu^{d}+\nu^{l})N \cdot 2z(y) ~dy\cdot v_x(y) \non\\
&=&2(\nu^{d}+\nu^{l})N \sqrt{(s_y)^2-y^2}~\dg y ~dy, \label{ydynoseki}
\eea
where we used 
$z(y)\!=\!\sqrt{(s_y)^2-y^2}$ and $v_x(y)\!=\!\dg y$. 
$\nu^{i}~(i\!=\!d,l)$ is the number of $i$-chains per unit volume.
Integrating (\ref{ydynoseki}) from $0$ to $\!s_y$
and multiplying by $2$ to take account of 
the region $y\!<\!0$, we have
the number of segments entering into the domain occupied by the 
loop in unit time:
\bea
2\!\int_0^{s_y}\!\!\!dy~\nu_{in}^{d,l}(y)
&=&(\nu^{d}+\nu^{l})N \frac{4}{3}(s_y)^3\dg \non\\ 
&=&\frac{4}{3 \cdot 36^{3/2}}(\nu^{d}+\nu^{l})a^3N^{5/2}\dg,
\label{ysekiunuti}
\eea
where we substituted $s_y$ given by (\ref{ynorads}) for the second equality.
Dividing (\ref{ysekiunuti}) by $u(\dg)$, we obtain  
\bea
A=\frac{1}{162}(\nu^{d}+\nu^{l})a^3N^{5/2}\frac{\dg}{u(\dg)}.
\label{Anohyousiki}
\eea
Putting (\ref{Anohyousiki}) into (\ref{konkai}), and solving it for
$u(\dg)$, we finally get
\bea
u(\dg)/\beta_0=
\frac{1\!+\!\sqrt{1\!+\!\frac{1}{162}(\Phi^d\!+\!\Phi^l) 
N^{3/2}\dg/\beta_0}}{2},
\label{tempusiki}
\eea
where $\Phi^i\!\equiv\!N\nu^ia^3$ is the volume fraction of $i$-chains.
In the experiments we are going to analyze,
$\Phi^a$ is much smaller than $\Phi^d$ and $\Phi^l$, as will be shown later.
In this case,  
$\Phi^d + \Phi^\ell$ in (\ref{tempusiki}) can be replaced by the volume 
fraction $\Phi$ of all chains.
Note that there is no $\br$ dependence in 
$u(\dot{\gamma})$, since the end-to-end length of the loop is zero.

\section{Observable Quantities}
\label{analy}

\subsection{Zero-Shear Viscosity}

Here, we derive the zero-shear viscosity $\eta_0$. 
Since we treat the case in which 
the shear rate $\dg$ is much smaller than $\beta_0$,
$\beta(r)$ can be regarded as $\beta_0$.
We use the basic equations (\ref{ware33ff}) 
written in terms of the Eulerian description.
Substituting (\ref{activevnaba}), (\ref{dkore1}) and (\ref{lkore0})
into (\ref{ware33ff}), we have a set of equations for 
$\phi_f^a(\br,t)$, $\phi_f^d(\br,t)$ and $\phi_f^l(\br,t)$:
\bea
& &\frac{\partial}{\partial t}\phi_f^a(\br,t)
+\nabla
\!\cdot\!\Biggl( \Biggl[\hat{\kappa}(t)\br\Biggr]\phi_f^a(\br,t) \Biggr) \non\\
& &\hspace*{2cm}
=-\beta_0\phi_f^a(\br,t)+p\phi_f^d(\br,t), \label{loeusyo00} \\
& &\frac{\partial}{\partial t}\phi_f^d(\br,t)+\nabla
\!\cdot\! \Biggl(\Biggl[\Biggl(
 \hat{\kappa}(t)\!-\!\frac{K_N}{\zeta_N}\Biggl)\br\!+\!\frac{1}{\zeta_N}\bR(t) 
\Biggr] \phi_f^d(\br,t) \Biggr) \non\\
& &\hspace*{1cm}=\beta_0\phi_f^a(\br,t)-(p+v(\br))\phi_f^d(\br,t)+
u\phi_f^l(\br,t),\label{dsli} \\
& &\frac{\partial}{\partial t}\phi_f^l(\br,t)
=v(\br)\phi_f^d(\br,t)-u\phi_f^l(\br,t).\label{loeusyo}
\eea
Taking average with respect to the stochastic process $\bR(t)$ in above 
equations, we obtain 
\bea
& &\frac{\partial}{\partial t}\phi^a(\br,t)
+\nabla
\!\cdot\!
\Biggl( \Biggl[\hat{\kappa}(t)\br\Biggr] \phi^a(\br,t) \Biggr) \non\\
& &\hspace*{2cm}=-\beta_0\phi^a(\br,t)+p\phi^d(\br,t), \label{loeusyof01} \\
& &\frac{\partial}{\partial t}\phi^d(\br,t)
+\nabla
\!\cdot\! \Biggl(\Biggl[\Biggl(
\hat{\kappa}(t)\!-\!\frac{K_N}{\zeta_N}\Biggl)\br\!-\!\frac{k_BT}{\zeta_N}
\nabla\Biggr] \phi^d(\br,t) \Biggr) \non\\
& &\hspace*{1cm}
=\beta_0\phi^a(\br,t)-(p+v(\br))\phi^d(\br,t)+
u\phi^l(\br,t),\label{dji2}\\
& &\frac{\partial}{\partial t}\phi^l(\br,t)
=v(\br)\phi^d(\br,t)-u\phi^l(\br,t)\label{ljfk}.
\eea
We will give their derivations in appendix \ref{sttt2}.
The equation (\ref{dsli}) corresponds to
the {\it stochastic Liouville equation},~\cite{kubo}
whereas the equation (\ref{dji2}) is equivalent to the 
{\it Fokker-Planck equation}.\footnote{The 
stochastic Liouville equation is the one for the probability 
distribution function in phase space where a point in the space 
represents a state of the system.
Since the mathematical structure of (\ref{dsli}) is the same as 
that of the stochastic Liouville equation,
we will call (\ref{dsli}) the stochastic Liouville equation.
In the same sense, we will call (\ref{dji2}) the Fokker-Planck equation.
}

In the long time limit $t\!\to\!\infty$, 
$\phi^i(\br,t)$ becomes independent of time (we denote it as $\phi^i(\br)$). 
Letting $\frac{\partial}{\partial t}\phi^i(\br,t)\!=\!0~(i\!=\!a,d,l)$ 
in the above equations, we have
\bea
& &
\nabla\!\cdot\!\Biggl(\hat{\kappa}\br  
\phi^a(\br) \Biggr)=-\beta_0\phi^a(\br)+p\phi^d(\br), \label{ap1}\\
& &
\nabla\!\cdot\! \Biggl(\Biggl[\Biggl(
 \hat{\kappa}\!-\!\frac{K_N}{\zeta_N}\Biggl)\br\!-\!\frac{k_BT}{\zeta_N}
\nabla
\Biggr] \phi^d(\br) \Biggr) \non\\
& &\hspace{1cm}=\beta_0\phi^a(\br)-(p+v(\br))\phi^d(\br)+
u\phi^l(\br),\label{dpp21}\\
& &v(\br)\phi^d(\br)=u\phi^l(\br)\label{lp1}.
\eea
With the help of (\ref{lp1}), we can eliminate $\phi^l(\br)$ in 
(\ref{dpp21}):\footnote{
(\ref{dp1}) is the equation used by Vaccaro and Marrucci~\cite{Mar1}
in the long time limit.}
\bea
& &
\nabla\!\cdot\! \Biggl(\Biggl[\Biggl(
\hat{\kappa}\!-\!\frac{K_N}{\zeta_N}\Biggl)\br\!-\!\frac{k_BT}{\zeta_N}
\nabla\Biggr] \phi^d(\br) \Biggr)  \non\\
& & \hspace*{3cm}
=\beta_0\phi^a(\br)-p\phi^d(\br). \label{dp1}
\eea

\subsubsection{Number of Chains in the Absence of Flow}
\label{numtaur}

The number of $i$-chains per unit volume in real space is
given by
\bea
\nu^{i}=\int d\br ~\phi^i(\br).\label{nuiteig}
\eea
According to (\ref{warerrerer22}), we obtain
the equation for the number-conservation
\bea
n=\nu^a+\nu^d+\nu^l,\label{nhozon}
\eea
where $n$ is the total number of chains in unit volume: 
\bea
n=\int d\br ~\phi (\br).
\eea

In the current situation of the slow shear flow,
$\nu^i$ is expanded in powers of 
$\dg$:~$\nu^i\!=\!\nu^i_0\!+\!\nu^i_1|\dg|\!+\cdots$.
Since the zeroth order term $\nu_0^i$ contributes
to $\eta_0$ as seen later, 
we now derive the expression for $\nu^i_0$, i.e.,
the number of $i$-chains in unit volume
at the equilibrium state.
Integrating (\ref{ap1}) or (\ref{dp1}) for $\hat{\kappa}\!=\!0$ , we have
\bea
0=-\beta_0 \nu^a_0+p\nu^d_0. \label{ap2}
\eea
Integrating (\ref{lp1}) for $\hat{\kappa}\!=\!0$, we obtain
\bea
v_0V\phi^d_0(\br=0)=\beta_0\nu^l_0,\label{dp22}
\eea
where $\phi^i_0(\br)$ means $\phi^i(\br)$ 
at the equilibrium state, 
and we put $u\!=\!\beta_0$ since 
$\dg\!=\!0$.
Let us now introduce $\psi(\br)$ by 
\bea
\phi^d_0(\br)=\nu^d_0\psi(\br). \label{psiohatu}
\eea
Note that because of (\ref{nuiteig}), $\psi(\br)$ is normalized as
$\int \!d\br \psi (\br)\!=\!1$.
Defining $V$ by
\bea
V=\frac{1}{\psi(\br=0)}, \label{defVV}
\eea
(\ref{dp22}) reduces to
\bea
v_0\nu^d_0=\beta_0\nu^l_0.\label{dp2}
\eea
Solving (\ref{ap2}) and (\ref{dp2}) with the help of (\ref{nhozon}),
we obtain 
\bea
\nu^{a}_0&=&\frac{n}{1+\frac{\beta_0}{p}(1+\frac{v_0}{\beta_0})},
\label{nua0notoki}\\
\nu^{d}_0&=&\frac{n}{1+\frac{p}{\beta_0}+\frac{v_0}{\beta_0}},
\label{nud0notoki} \\
\nu^{l}_0&=&\frac{n}{1+\frac{\beta_0}{v_0}(1+\frac{p}{\beta_0})}.
\label{lud0notoki}
\eea

\subsubsection{Root-Mean-Square of End-to-End Vector}

Hereafter, we denote the expectation value of 
the physical quantity $A(\br)$ with respect to the $i$-chain as
\bea
\bra A(\br)\ket^i\!=\!\frac{\int \!d\br A(\br)\phi^i(\br)}
{\int \!d\br\phi^i(\br)} 
\!=\!\frac{1}{\nu^i}\!\int \!d\br A(\br)\phi^i(\br).\label{anokitaiti}
\eea  
Multiplying the equation (\ref{ap1}) and (\ref{dp1}) by 
$r_{\alpha}r_{\beta}$,
and integrating them with respect to $\br$, we have
\bea
& &-\kappa_{\alpha\gamma}\bra r_{\gamma}r_{\beta}\ket^a\nu^a
-\bra r_{\alpha}r_{\gamma}\ket^a\kappa_{\beta\gamma}\nu^a \non\\
& &\hspace*{2cm}=-\beta_0\bra r_{\alpha}r_{\beta}\ket^a\nu^a
+p\bra r_{\alpha}r_{\beta}\ket^d\nu^d, \label{kanom1}\\
& &-\kappa_{\alpha\gamma}\bra r_{\gamma}r_{\beta}\ket^d\nu^d
-\bra r_{\alpha}r_{\gamma}\kappa_{\beta\gamma}\ket^d\nu^d
+\frac{2K_N}{\zeta_N}\bra r_{\alpha}r_{\beta}\ket^d\nu^d \non\\
& &-\frac{2k_BT}{\zeta_N}\delta_{\alpha\beta}\nu^d 
=\beta_0\bra r_{\alpha}r_{\beta}\ket^a\nu^a
-p\bra r_{\alpha}r_{\beta}\ket^d \nu^d.\label{kanom2}
\eea
Substituting (\ref{shearlamkap}) for $\hat{\kappa}$, we obtain
\bea
& &\hspace*{-0.5cm}
-2\dg\bra xy\ket^a\nu^a=-\beta_0\bra x^2\ket^a\nu^a+p\bra x^2\ket^d\nu^d, \\
& &\hspace*{-0.5cm}
-\dg\bra y^2\ket^a\nu^a=-\beta_0\bra xy\ket^a\nu^a+p\bra xy\ket^d\nu^d,\\
& &\hspace*{-0.5cm}0=-\beta_0\bra y^2\ket^a\nu^a+p\bra y^2\ket^d\nu^d,\\
& &\hspace*{-0.5cm}
-\frac{2k_BT}{\zeta_N}\nu^d+\frac{2K_N}{\zeta_N}\bra xy\ket^d\nu^d
-2\dg\bra xy\ket^d\nu^d \non\\
& &\hspace*{3cm}=\beta_0\bra x^2\ket^a\nu^a-p\bra x^2\ket^d\nu^d, \\
& &\hspace*{-0.5cm}
\frac{2K_N}{\zeta_N}\bra xy \ket^d\nu^d\!-\!\dg\bra y^2\ket^d\nu^d
\!=\!\beta_0\bra xy\ket^a\nu^a\!-\!p\bra xy\ket^d\nu^d, \\
& &\hspace*{-0.5cm}
-\frac{2k_BT}{\zeta_N}\!+\!\frac{2K_N}{\zeta_N}\bra y^2\ket^d\nu^d
\!=\!\beta_0\bra y^2\ket^a\nu^a\!-\!p\bra y^2\ket^d\nu^d,
\eea
where we put $x\!=\!r_1$, $y\!=\!r_2$ and $z\!=\!r_3$.
Solving the above equations,
we get root-mean-square 
of the end-to-end vector for 
the active and the dangling chain in the form
\bea
& &\hspace*{-0.5cm}\bra x^2 \ket^a=\frac{Na^2}{3} 
\Biggl[ 1+2\Biggl( \frac{\dg}{\beta_0}\Biggr)^{\hspace*{-0.1cm}2}\non\\
& &\hspace*{0.7cm}+2(\dg \tau)^2 
\Biggl\{\!\Biggl(\! 1\!+\!\frac{p}{\beta_0} \!\Biggr)^{\hspace*{-0.1cm}2}
\!+\!\frac{1}{\beta_0\tau}\Biggl(\!1\!+\!\frac{2p}{\beta_0} 
\!\Biggr)^{\hspace*{-0.1cm}2} \Biggr\} \Biggr], \label{x2asecon}\\
& &\hspace*{-0.5cm}\bra xy \ket^a=\frac{Na^2}{3}
\Biggl[\frac{\dg}{\beta_0}+\dg\tau\Biggl( 1+\frac{p}{\beta_0} \Biggr)\Biggr],
\label{xyasecon}\\
& &\hspace*{-0.5cm}\bra y^2 \ket^a=\frac{Na^2}{3},\label{y2asecon}\\
& &\hspace*{-0.5cm}\bra x^2 \ket^d= \frac{Na^2}{3}
\Biggl[ 1\!+\!2(\dg \tau)^2\Biggl\{
\Biggl( 1\!+\!\frac{p}{\beta_0} \Biggr)^{\hspace*{-0.1cm}2}\!\!+
\!\frac{p}{\beta_0^2\tau}\Biggl\}\Biggr],
\label{dselx2}\\
& &\hspace*{-0.5cm}
\bra xy \ket^d=\frac{Na^2}{3}\dg\tau\Biggl( 1+\frac{p}{\beta_0} \Biggr),
\label{dselxy}\\
& &\hspace*{-0.5cm}\bra y^2 \ket^d=\frac{Na^2}{3},\label{dsely2}\\
& &
\Biggl(~\bra z^2 \ket^a=\bra z^2 \ket^d=\frac{Na^2}{3},
~~\mbox{(others)=0} \Biggr)
\label{nokorim}
\eea
where $\tau$ is given by (\ref{deftau}).

\subsubsection{Zero-Shear Viscosity}

The shear stress caused by $i$-chains is given by
\bea
\sigma_{xy}^i
&=&\nu^i\bra \frac{xy}{r}f(r) \ket^i . \label{defshest}
\eea
We can use the linear form for the tension $f(r)$ of the chain,
since, when $\dg\ll\beta_0$, the end-to-end length $r$ of the chain
stays sufficiently short during the deformation 
compared with its contour length.
Putting $f(r)\!=\!K_Nr$ into (\ref{defshest}), we have
\bea
\sigma_{xy}^i
&=&K_N\nu^i\bra xy\ket^i.
\eea
With the help of (\ref{xyasecon}), we obtain the shear stress caused by
active chains in the form
\bea
\sigma_{xy}^a
&=&k_BT \nu^a
\Biggl[\frac{\dg}{\beta_0}+\dg\tau\Biggl( 1+\frac{p}{\beta_0} \Biggr)\Biggr]
\non\\
&=&k_BT \nu^a_0
\Biggl[\frac{\dg}{\beta_0}+\dg\tau\Biggl( 1+\frac{p}{\beta_0} \Biggr)\Biggr]
+O(\dg^2),\label{xyasecon2}
\eea
where $\nu^a_0$ is given by (\ref{nua0notoki}).
Then, the zero-shear viscosity is obtained as
\bea
\eta^a_0=\lim_{\dg\to0}\frac{\sigma_{xy}^a}{\dg}=k_BT \nu^a_0
\Biggl[\frac{1}{\beta_0}+\tau\Biggl( 1+\frac{p}{\beta_0} \Biggr)\Biggr].
\label{etaa0}
\eea
Similarly, with the help of (\ref{dselxy}), 
the shear stress caused by dangling chains becomes
\bea
\sigma_{xy}^d
&=&k_BT \nu^d_0\dg\tau\Biggl( 1+\frac{p}{\beta_0} \Biggr)+O(\dg^2), 
\label{zerosigma}
\eea
where $\nu^d_0$ is given by (\ref{nud0notoki}), and
the zero-shear viscosity is obtained as	
\bea
\eta^d_0=\lim_{\dg\to0}\frac{\sigma_{xy}^d}{\dg}
=k_BT \nu^d_0\tau\Biggl( 1+\frac{p}{\beta_0} \Biggr). \label{etan0}
\eea
The total zero-shear viscosity is given by $\eta_0=\eta_0^a+\eta_0^d$.

\subsubsection{Remarks on the Relaxation Time}
\label{kanwatime}

In the experiments we are going to analyze,\cite{Jenkins1} 
the relaxation time 
$\tau$ ($\simeq\!10^{-5}$s, see the next section for its estimation) 
given by (\ref{deftau}) 
is always much shorter than the specific time 
$1/\dg (=\!10^{-3}\!\sim10^{-1}$s) of the shear flow. 
In this case, as we can see from (\ref{xyasecon2}) and 
(\ref{zerosigma}), the effect of $\tau$ becomes very week
for $\dg$ smaller than $1/\tau(\simeq\!10^5$s). 
Consequently, we will adopt the limit $\tau\!\to\!0$ from now on.
In this limit, the dangling chain has always stress-free conformation
which is regarded as the Gaussian.\footnote{
This limit was adopted in some literatures.~\cite{TE0,Wang}}
Then $\phi^d(\br)$ is written, even for finite $\dg$ (see (\ref{psiohatu})), 
as
\bea
\phi^d(\br)=\nu^d\psi(\br),
\eea
where 
\bea
\psi(\br)=\Biggl( \frac{3}{2\pi Na^2}\Biggr)^{\hspace*{-0.1cm}{3/2}}
\!\exp\Biggl[-\frac{3r^2}{2Na^2}\Biggr]. \label{gausspsi}
\eea
The total zero-shear viscosity becomes at $\tau\!\to\!0$
\bea
\eta_0=\nu^a_0\frac{k_BT}{\beta_0} 
=\frac{n}{1+
\frac{\beta_0}{p}(1+\frac{v_0}{\beta_0})}\frac{k_BT}{\beta_0}. \label{eta0}
\eea

\subsection{Non-Newtonian Viscosity}

Here, we investigate the shear viscosity
for the shear rate including larger than $1/\beta_0$.
To this end, we analyze the basic equations (\ref{lagsekibun}) 
written in terms of the Lagrangian description,\footnote{
Note that we can use (\ref{lagsekibun}) instead of (\ref{lagsekibunff})
when $\tau$ is 0.} 
since this description connects $\phi^a(\br,t)$ (and $\phi^l(\br,t)$) 
with $\psi(\br)$.

\subsubsection{Equation for Active Chains}

The equation for the active chains is given by putting $i\!=\!a$ in 
(\ref{lagsekibun}) as
\bea
& &J^a(t,t_0)\phi^a(\br^a(t;\br_0,t_0),t) \non\\
& &=\exp\Biggl[ -\int_{t_0}^tdt^{\prime}~\beta(\br^a(t^{\prime};\br_0,t_0))
\Biggr]\phi^a(\br_0,t_0) \non\\
& &\hspace*{0.3cm}+p\int_{t_0}^tdt^{\prime}~
\exp\Biggl[-\int_{t^{\prime}}^tdt^{\prime\prime}
\beta(\br^a(t^{\prime\prime};\br_0,t_0))\Biggr] \non\\
& &\hspace*{1.4cm}\times J^a(t^{\prime},t_0)\nu^d(t^{\prime})
\psi(\br^a(t^{\prime};\br_0,t_0)),\label{acrare}
\eea
where we used the relation
\bea
\phi^d(\br,t)=\nu^d(t)\psi(\br).
\eea
$\psi(\br)$ is given by (\ref{gausspsi}).
Recall that 
$\br^a(t;\br_0,t_0)$ in (\ref{acrare}) is given 
by (\ref{tnoto3}).
In order to obtain the equation for the expectation value of $A(\br)$ 
with respect to the active chain,
let us multiply the equation (\ref{acrare}) by $A(\br^a(t;\br_0,t_0))$
and integrate it with respect to $\br_0$: 
\bea
& &\hspace*{-0.1cm}\int d\br~A(\br)\phi^a(\br,t) \non\\
& &\hspace*{-0.1cm}=\!\!\int \!\!d\br
\exp\Biggl[ \!-\!\int_{t_0}^t\!dt^{\prime}\beta(\br^a(t^{\prime};\br,t_0))
\Biggr]A(\br^a(t;\br,t_0))\phi^a(\br,t_0) \non\\
& &\hspace*{0.2cm}+p\!\int_{t_0}^t\!dt^{\prime}\!\int 
d\br \exp\Biggl[-\!\int_{t^{\prime}}^tdt^{\prime\prime}
\beta(\br^a(t^{\prime\prime};\br,t^{\prime}))\Biggr] \non\\
& &\hspace*{1cm}\times
A(\br^a(t;\br,t^{\prime}))\nu^d(t^{\prime})\psi(\br).\label{actotyu}
\eea
In deriving (\ref{actotyu}), we changed 
the integration variable from $\br^a(t;\br_0,t_0)$ to $\br$ in
the left-hand side of (\ref{actotyu}), 
and from $\br^a(t^{\prime};\br_0,t_0)$ 
to $\br$ in the second term of the right-hand side.
Under the steady flow, due to
$\lambda(t;t^{\prime})\!=\!\lambda(t-t^{\prime})$, 
(\ref{actotyu}) reduces to
\bea
& &\hspace*{-0.3cm}\int d\br~A(\br)\phi^a(\br,t) \non\\
& &\hspace*{-0.3cm}=\int \!\!d\br
\exp\Biggl[ \!-\!\int_{0}^{t-t_0}\hspace*{-0.5cm}
dt^{\prime}\beta(\br^a(t^{\prime};\br))
\Biggr]A(\br^a(t-t_0;\br))\phi^a(\br,t_0) \non\\
& &\hspace*{-0cm}+p\int_{0}^{t-t_0}\hspace*{-0.3cm} dt^{\prime}\!\int 
d\br\exp\Biggl[-\!\int_{0}^{t^{\prime}} \!\!dt^{\prime\prime}
\beta(\br^a(t^{\prime\prime};\br))\Biggr] \non\\
& &\hspace*{1cm}\times A(\br^a(t^{\prime};\br))\nu^d(t-t^{\prime})
\psi(\br), \non\\ \label{tlimnototyu}	
&=&\int_{D_{r^*}^{\prime}(t-t_0)} \hspace*{-1cm}d\br~
\e^{-\beta_0(t-t_0)}A(\br^a(t-t_0;\br))\phi^a(\br,t_0) \non\\
&+&p\!\int_{0}^{t-t_0}\hspace*{-0.5cm}dt^{\prime}\!
\int_{D_{r^*}^{\prime}(t^{\prime})} \hspace*{-0.6cm}
d\br~\e^{-\beta_0t^{\prime}}A(\br^a(t^{\prime};\br))\nu^d(t-t^{\prime})
\psi(\br),  \label{tlimnototyu2}	
\eea
where 
\bea
\br^a(t;\br)=\hat{\lambda}(t)\br \label{ranokoko}
\eea
and
\bea
D_{r^*}^{\prime}(t)&=&\{ \br~|~r^a(t^{\prime};\br)\le r^*~ 
(\forall t^{\prime}\le t) \}.
\eea

A steady state can be obtained by taking 
the limit $t\!\to\!\infty$ and $t_0\!\to\!-\infty$.
Clearly, the first term in the right-hand side of (\ref{tlimnototyu2})
becomes 0 in these limit.
As for the second term, because of $t\!\to\!\infty$,
$\nu^d(t-t^{\prime})$ can be regarded as the value 
$\nu^d$ in the steady state for
$t^{\prime}\!<\!\infty$. 
On the other hand, for 
$t^{\prime}\!\to\!\infty$, 
the integrand
becomes exponentially small due to the factor $\e^{-\beta_0t^{\prime}}$.
Therefore, $\nu^d(t-t^{\prime})$ is always regarded as $\nu^d$
in the limit of steady state.
Hence, 
(\ref{anokitaiti}) for $i\!=\!a$ becomes
\bea
\hspace*{-0.6cm}
\bra A(\br) \ket^a\!=
p\frac{\nu^d}{\nu^a}\!\!
\int_{0}^{\infty}\hspace*{-0.3cm}
dt \! \int_{D_{r^*}^{\prime}(t)} \hspace*{-0.6cm}
d\br~\e^{-\beta_0 t}
A(\br^a(t;\br))\psi(\br).\label{actlimDp}
\eea
Hereafter, 
we will take 
\bea
D_{r^*}(t)=\{ \br~|~r^a(t;\br) \le r^*\}
\eea
instead of $D_{r^*}^{\prime}(t)$ as
the region of the integration in (\ref{actlimDp})
as far as the steady shear flow is concerned. 
Then the contribution to the integration (\ref{actlimDp}) from
the region outside $D_{r^*}^{\prime}(t)$ 
(i.e., $r\!=\!r^a(t\!=\!0;\br)\!>\!r^*$) 
emerges. 
Recall that the Gaussian distribution function $\psi(\br)$
in the integrand of (\ref{actlimDp}) 
mainly weighs the contribution from $r$ smaller than $\sqrt{N}a$.
Since it turns out that $r^*$ is close to $Na$, 
the contribution 
from the region $r\!>\!r^*
(\simeq\! Na\!\gg\!\sqrt{N}a)$ is almost negligible.
Therefore, we can 
substitute safely the region $D_{r^*}(t)$ for $D_{r^*}^{\prime}(t)$.

\subsubsection{Equation for Loops}
\label{analopsub}

The equation for the loops is written as
\bea
& &\hspace*{-1.2cm}J^l(t,t_0)\phi^l(\br^l(t;t_0,\br_0),t)
=\e^{-u(t-t_0)}\phi_f^l(\br_0,t_0) \non\\
& &\hspace*{0cm}+\int_{t_0}^t\!\!dt^{\prime}
\e^{-u(t-t^{\prime})}
J^l(t^{\prime},t_0)\nu^d(t^{\prime}) \non\\ 
& &\hspace*{1cm}\times
v(\br^l(t^{\prime};t_0,\br_0))
\psi(\br^l(t^{\prime};t_0,\br_0),t^{\prime}).
\eea
Following the same procedure as the case of active chains,
we have the equation for the expectation value of $A(\br)$ 
with respect to the loop:
\bea
\int \! d\br A(\br)\phi^l(\br)&=&\frac{\nu^d}{\nu^l}
\int_{0}^{\infty}\!\!\!dt \! \int \!d\br~\e^{-ut}
v(\br)A(\br^l(t;\br))\psi(\br)\non\\
&=&\frac{\nu^d}{\nu^l}\frac{v_0}{u}V\psi(0)A(0) \non\\
&=&\frac{\nu^d}{\nu^l}\frac{v_0}{u}A(0), \label{looplim}
\eea
where we used $V$ given by (\ref{defVV}).

\subsubsection{Number of Chains}

Putting $A(\br)=1$ in (\ref{actlimDp}),
we have 
\bea
\nu^a=p\nu^d\!\int_{0}^{\infty}\!\!dt \! \int_{D_{r^*}(t)} 
\hspace*{-0.5cm}d\br~
\e^{-\beta_0t}\psi(\br)=\nu^d\frac{p}{\zeta},
\eea
where we introduced
\bea
\frac{1}{\zeta}=
\int_0^{\infty}\!dt \!\int_{D_{r^*}(t)} \hspace*{-0.5cm}
d\br~\e^{-\beta_0t}\psi(\br).\label{nuanomo}
\eea
Note that when $\dg\!\to\!0$, we find $\zeta\!\to\!\beta_0$.
Similarly, by putting $A(\br)\!=\!1$ in (\ref{looplim}), we have
\bea
\nu^l=\nu^d\frac{v_0}{u}. \label{nulnomo}
\eea
By solving (\ref{nuanomo}) and (\ref{nulnomo}) with the help of (\ref{nhozon}),
the numbers of chains per volume for each type becomes
\bea
\nu^{a}&=&\frac{n}{1+\frac{\zeta}{p}(1+\frac{v_0}{u})},\label{audzeta}\\
\nu^{d}&=&\frac{n}{1+\frac{p}{\zeta}+\frac{v_0}{u}}, \label{nudzeta}\\
\nu^{l}&=&\frac{n}{1+\frac{u}{v_0}(1+\frac{p}{\zeta})}.  \label{ludzeta}
\eea

\subsubsection{Shear Viscosity}

The shear stress caused by active chains is 
\bea
\sigma_{xy}^a
&=&\nu^a\bra \frac{xy}{r}f(r) \ket^a \non\\
&=&p\nu^d\frac{k_BT}{a}\int_0^{\infty} dt 
\int_{D_{r^*}(t)} \hspace*{-0.5cm}d\br ~\e^{-\beta_0t} \non\\
& &\times\frac{x^a(t;\br)y^a(t;\br)}{r^a(t;\br)}L^{-1}(r^a(t;\br))
\psi(\br), \label{saigonouryo}
\eea
where $\nu^d$ is given by (\ref{nudzeta}) and $\br^a(t,\br)$
is given by (\ref{ranokoko}).
Recalling that the shear stress caused by dangling chains is zero, 
we finally get the 
formula of the total shear viscosity in the form
\bea
\eta=\frac{\sigma_{xy}^a}{\dg}.\label{visfinal}
\eea

\subsection{Tanaka-Edwards Limit}
\label{secTE}

Tanaka and Edwards~\cite{TE1} 
calculated the steady shear viscosity for 
the physical gel, composed of active and dangling chains (loops are absent),
with the assumptions that 
1) the breakage rate is given by $\beta(r)\!=\!\beta_0\!+\!\beta_1r^2$
($\beta_0$ and $\beta_1$ are constant), and
2) even the active chains can be described by Gaussian chain model
with the Hookean force $f_N^0(r)\!=\!K_Nr$.
The assumption 1) indicates that 
there are finite populations of active chains nearly equal to and/or 
even longer than the contour length $l$, i.e., there is
no cutoff length $r^*$ $(<\!l)$ in the TE model.
The Gaussian chain model, proposed in the assumption 2), 
means that there are only shorter active chains ($r\!\ll\!l$) 
satisfying the Hookean force assumption. 
In this sense, the above two assumptions contradict each other.

This conflict between the two assumptions 1) and 2) can be resolved by
introducing the cutoff length $r^*$ much smaller than $l$.
However, in general, there seems no reason to restrict the force to be linear.
That is the reason why we adopt the force (\ref{langtyoury}) given by the 
random-flight model with the cutoff $r^*$ ($0\!<\!r^*\!<\!l$). 
Hereafter, we call the case where $r^*\!\ll\!l$ and $v_0\!\to\!0$
the {\it TE limit} within our model, since, in this case,
the Hookean force is valid and loops are absent ($\nu^l\!=\!0$).


\section{Analyses of Experiments}
\label{results}

Now, we analyze, with the help of the formulae 
(\ref{visfinal}) and (\ref{saigonouryo}) derived theoretically in this paper, 
the steady shear viscosity observed experimentally by 
Jenkins {\it et al.}~\cite{Jenkins1} for HEUR having hexadecyl 
end groups in water. 

\subsection{Determination of $N$, $a$ and $\Phi$} 
\label{nphisec}

\begin{table*}[t]
\begin{center}
\begin{tabular}{c|c|c|c|c|c|c} 
$M_n$ & $n_{rep}$ & $l$ (\AA) 
& $\sqrt{\bra r^2\ket_0}$ (\AA) 
& $N$ & $a$(\AA) & $\Phi$ \\ 
      & $=\!M_n/M_0$ & $=\!l_0n_{rep}$ & 
& $=\!l^2/\bra r^2\ket_0$ & $=\!l/N$  & $=\!10^4cN_ANa^3/M_n$  \\ \hline
34,200 & 777   & 2,800  & 186 &  227   & 12.3 & 7.4$\times10^{-2}$ \\
51,000 & 1,160 & 4,180  & 231 &  327   & 12.8 & 8.1$\times10^{-2}$ \\
67,600 & 1,540 & 5,540  & 265 &  437   & 12.7 & 8.0$\times10^{-2}$ \\
84,300 & 1,920 & 6,910  & 301 &  527   & 13.1 & 8.5$\times10^{-2}$ \\ 
\end{tabular}
\end{center}
\caption{Properties of HEUR in water. 
$M_n$:~number-average molecular weight,~\cite{Jenkins1,Jenkins5}
$n_{rep}$:~number of the EO units, $l$:~contour length,
$\protect\sqrt{\protect\bra r^2\protect\ket_0}$:~root-mean-square 
of the end-to-end length,~\protect\cite{Jenkins5} 
$N$:~number of segments, $a$:~length of each segment 
and $\protect\Phi$:~volume fraction 
for $c\protect\!=\protect\!1$wt\protect\% solution
(mean value$=\!8.0\!\times\!10^{-2}$).
$N_A\!=\!6.02\!\times\!10^{-23}$ is the Avogadro constant.
For EO, $M_0\!=\!44$ and $l_0\!=\!3.6$\protect\AA.
}
\label{hyoup2}
\end{table*}

The contour length $l$ of HEUR is obtained if we know the 
fully extended length $l_0$ of ethylene oxide (EO), the main component of 
HEUR, and the number $n_{rep}$ of EO units in HEUR, i.e., 
$l\!=\!l_0\!\times\!n_{rep}$.
Taking the value 1.54\AA\ for a carbon-carbon bond length, 1.43\AA\ for 
a carbon-oxygen bond length, and 70.5$^\circ$ for the angle formed by 
two bonds (tetrahedral bonds) in polymer backbone, we have 
$l_0\!=\!(1.54\!+\!2\!\times\!1.43)\!\times\!\cos(70.5^{\circ}/2)\!=\!3.6$\AA.
Letting $M_0(=\!44$) stands for the molecular weight of the EO unit,
we have $n_{rep}\!=\!M_n/M_0$, where $M_n$ is the molecular weight of HEUR.
The values of $l$ are listed in the third column of Table~\ref{hyoup2}
for each $M_n$ of the experiment.~\cite{Jenkins1}

Knowing the values of the root-mean-square $\sqrt{\bra r^2 \ket_0}$ 
of the end-to-end 
length\footnote{In the literature,~\cite{Jenkins5} 
four distinct values of $\sqrt{\bra r^2\ket_0}$ at $30^\circ$C
are reported for each molecular weight, 
which were obtained using four different theoretical relations
connecting $\sqrt{\bra r^2\ket_0}$ and the intrinsic viscosity $[\eta]$.
The values of $\sqrt{\bra r^2\ket_0}$ 
listed in Table~\ref{hyoup2} are mean values of these data.
Although the steady shear viscosity we are going to analyze is
observed at $25^\circ$C, since $[\eta]$, and therefore 
$\sqrt{\bra r^2\ket_0}$, do not change their values so much
within the temperature range $25\!\sim\!30^\circ$C (see
Figures 7,8 and 9 in the literature~\cite{Jenkins5}),
we can use $\sqrt{\bra r^2\ket_0}$ at $30^\circ$C to analyze
the steady shear viscosity at $25^\circ$C.}
in addition to $l$, we can obtain the number 
$N$ of segments per chain and the length $a$ of each segment by the relations 
$N\!=\!l^2/\bra r^2 \ket_0$ and $a\!=\!l/N$.
Here, $\bra\cdots\ket_0$ indicates the average taken 
in equilibrium state without any flow.
The values of $N$ and $a$ are 
listed respectively in the fifth and sixth 
column of Table~\ref{hyoup2}.
We see that the length $a$ of the segment is almost independent of 
molecular weight $M_n$ (or $N$).
Since the fully extended length $l_0$ of the EO unit is 3.6\AA,
we can conclude that one segment contains $3\!\sim\!4$ units.

The volume fraction of polymers is defined by 
$\Phi\!=\!Nna^3$ with $n$ being the number density of polymers. 
Since the number density is related to the mass density $\rho$ 
of polymers by $n\!=\!\rho N_A/M_n$, the volume fraction reduces to 
$\Phi\!=\!10^4 c N_A N a^3/M_n$ where $N_A$ is the Avogadro constant. 
Note that the mass density $\rho$ is related to the polymer concentration 
$c$ expressed by the weight percentage by $\rho\!=\!10^4\!\times\!c$ (g/m$^3$).
For each $M_n$, the volume fraction $\Phi$ 
of the $c\!=\!1$wt\% HEUR aqueous solution 
is listed in Table \ref{hyoup2}. 
They are almost the same as they should be.
We adopt the mean value $\Phi\!=\!8.0\!\times\!10^{-2}$ 
of them in the following.
Note that the overlap threshold of polymers, estimated as 
$\Phi^*\!\simeq\!Na^3/\bra r^2\ket_0^{3/2}\!\simeq\!N^{-1/2}$, 
is roughly $7\!\times\!10^{-2}$ when $N\!=\!227$, for instance. 
It indicates that the volume fraction of polymers in 
this system, $\Phi\!=\!8.0\!\times\!10^{-2}$, 
is on the order of the overlap threshold. 

\subsection{Estimate of the Relaxation Time}
\label{rmatan2}

We mentioned in the preceding section
that the relaxation time $\tau$ given by (\ref{deftau}) 
(or $\tau_1$ given by (\ref{raustau})) 
is much smaller than the specific time of the flow.
This is quantitatively shown for HEUR by
putting the values of $a$ and $N$ listed in Table \ref{hyoup2} 
into (\ref{raustau}) with
$T\!=\!298$K and $\eta_s\!=\!0.89\times10^{-3}$Pa$\cdot$s 
(the viscosity of pure water at $\!298$K, 1atm).
The obtained relaxation time for each molecular weight 
is shown in Table \ref{hyoup3}.
Note that they have the order of $10^{-5}$s.
On the other hand, as seen from Figure \ref{TEviss},  
the interesting phenomena occur at a characteristic time 
in the range $10^{-3}\!\sim\!10^{-1}$s.
This is the quantitative 
reason why we took the limit $\tau$ (or $\tau_1)\!\to\!0$ in 
the previous section.

\begin{table}[t]
\begin{center}
\begin{tabular}{c|c} 
$M_n$  &  $\tau_1$(s)                \\\hline 
34,200 &  $1.3\!\times\!10^{-5}$    \\
51,000 &  $3.1\!\times\!10^{-5}$    \\
67,600 &  $5.4\!\times\!10^{-5}$    \\
84,300 &  $8.6\!\times\!10^{-5}$    \\ 
\end{tabular}
\end{center}
\caption{Rouse relaxation time of HEUR obtained from (\protect\ref{raustau})}
\label{hyoup3}
\end{table}

\subsection{Adjustable Parameters}

In our model, the shear viscosity $\eta$ 
is given by (\ref{visfinal}) with (\ref{saigonouryo}).
We see that there are four parameters in (\ref{saigonouryo}), i.e.,
$\beta_0$, $p$, $v_0$ and $r^*$ ($u$ is represented by $\beta_0$).
Making use of the zero-shear viscosity obtained from the 
experiments.~\cite{Jenkins1}
we can eliminate one of three 
parameters $\beta_0$, $p$, $v_0$
through the relation (\ref{eta0}).
For instance, 
$p$ is represented by $\beta_0$ and $v_0$ as 
\bea
p=\frac{\beta_0+v_0}{nk_BT/\eta_0\beta_0-1},\label{kansanp}
\eea
or, written in terms of $\beta_0$ and $v_0/p$ as
\bea
p=\frac{\beta_0}{nk_BT/\eta_0\beta_0-1-(v_0/p)}.\label{kansanpv0}
\eea
Hereafter, we adopt $\beta_0$ and $v_0/p$ (and $r^*$)
as adjustable parameters.
The ratio $v_0/p$ is a measure representing which channels of the transitions
are favorable, i.e., one is to the 
active, and the other to the loop from the dangling chain.


\subsection{Shear Viscosity}

\begin{figure} 
\begin{center}
\includegraphics*[scale=0.6]{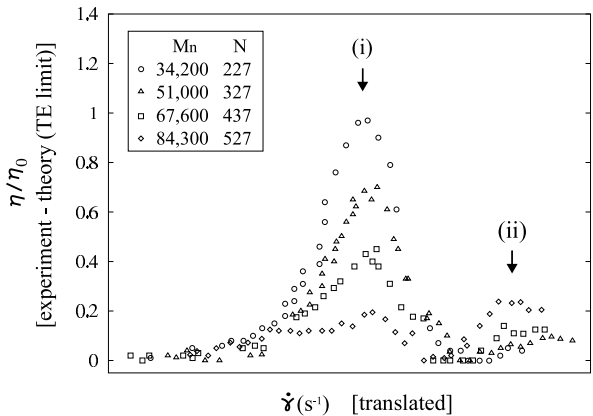}
\end{center}
\caption{The difference between theoretically (TE limit) and 
experimentally obtained shear viscosity.
The theoretical results 
(lines in Figure \protect\ref{TEviss} in appendix \protect\ref{appeTElimit})
are subtracted from experimental data 
(dots in Figure \protect\ref{TEviss})
for each molecular weight $M_n$ ($N$). 
To clarify the $M_n$ dependence of the height of peaks,
each data is translated so that the positions of the first peaks 
(i) coincide with each other.}
\label{TEgap}
\end{figure}

In order to clarify the problems to be resolved,
let us see the difference between 
the viscosity obtained from the TE limit 
(see Figure \ref{TEviss} in appendix \ref{appeTElimit}) 
and the experimentally observed one 
by subtracting the former from the latter for each molecular weight,
and by translating along the abscissa so that its position of the (first) 
peak coincides with others.
As can be seen in Figure \ref{TEgap}, there appear two peaks. 
The set of peaks at lower shear rates ((i), in Figure \ref{TEgap}) reflects
the shear-thickening behavior.
The height of the peaks 
becomes low with increasing $M_n$.
It seems that each peak consists of two elements, i.e., 
one is high and narrow with strong $M_n$-dependence, whereas the other is
low and broad with weak $M_n$-dependence.
In the set of peaks at higher shear rates, 
i.e., at shear-thinning region ((ii), in Figure \ref{TEgap}), 
the height of the peaks has opposite $M_n$ dependence compared to
the previous set of peaks.
We will study mainly in this paper the first peak in the following.

Let us here estimate $r^*$ from the viewpoint of
bonding energy among hydrocarbon end groups.
We consider the potential barrier with height $W$ 
around a junction.~\cite{TE1}
An active chain dissociates from the junction
when its one end climbs over the potential barrier $W$ 
by the force $\bfs(\br)$.
Since the effective range of the 
hydrophobic interaction, $\sigma$, is
sufficiently shorter than $r^*$, we can put $\sigma\!=\!a$.
Then, the cut-off length $r^*$ is roughly estimated by the condition
\bea
W=\int_{r^{\prime}=r^*}^{r^{\prime}=r^*-a}
d\br^{\prime}\cdot\bfs(\br^{\prime})\simeq f(r^*)a. \label{rstar}
\eea
The right-hand side of (\ref{rstar})
represents the work done by the tension $\bfs(\br)$ to detach the end 
from the junction when $r\!=\!r^*$.
Assuming that $r^*\!\lnear\!l$, we have from (\ref{langtyoury})
$W/k_BT\!\simeq\!(1\!-\!r^*/l)^{-1}$, therefore,
$r^*/l\!\simeq\!1\!-\!k_BT/W$.
According to Annable {\it et al.},~\cite{Annable1} 
$W\!=\!28k_BT$ ($T\!=\!298$K) for HEUR of $M_w\!\simeq\!35,000$
having hexadecyl end groups (called C16/35 in the literature).
In this case, $r^*$ is estimated as $r^*\!\simeq\!0.96l$.


\subsubsection{Effects of Loops}
\label{efflopp}

\begin{figure} 
\begin{center}
\includegraphics*[scale=0.6]{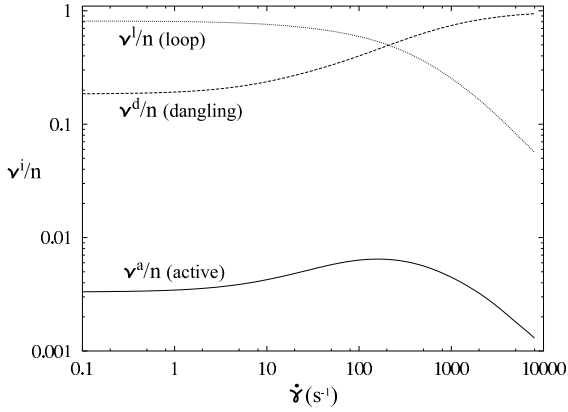}
\end{center}
\caption{The number of active chains, dangling chains 
and loops plotted against the shear rate 
for $r^*/l\!=\!0.97$, $\protect\beta_0\!=\!8$s$^{-1}$ and $v_0/p\!=\!245$
($N\!=\!227$) corresponding to Figure \protect\ref{Nloop22pv}.}
\label{Nloop22nupv0}
\end{figure}

The number 
of each type of chains are 
plotted in Figure \ref{Nloop22nupv0} for $r^*\!=\!0.97l$,
$\beta_0\!=\!8$s$^{-1}$ and $v_0/p\!=\!245$ ($N\!=\!227$).
With increasing the shear rate, the number $\nu^l$ of loops decreases
and the number $\nu^d$ of dangling chains increases, since
the transition probability $u$ from a loop to a dangling chain 
increases with shear rate (see (\ref{tempusiki})).
Due to the constant transition rate $p$ from a dangling to an active chain,
the number $\nu^a$ of active chains also increases with the same rate
as the dangling chains at lower shear rates.
That contributes to the gradual enhancement of the 
viscosity at lower shear rates as shown below.
The enhancement of the number of active chains 
at the shear-thickening region has been observed by 
Tam {\it et al.}.~\cite{Tam}
For higher shear rates, 
the population of active chains whose end-to-end length reaches $r^*$
becomes large, which causes the decrease in $\nu^a$ 
leading to the shear-thinning.
The numbers of active and dangling chains
and the number of loops in these figures indicate that
the assumption 
$\nu^a\!\ll\!n^d$ and $\nu^a\!\ll\!n^l$  
in (\ref{tempusiki}) is reasonable 
(it also holds for other values of $N$).

\begin{figure} 
\begin{center}
\includegraphics*[scale=0.6]{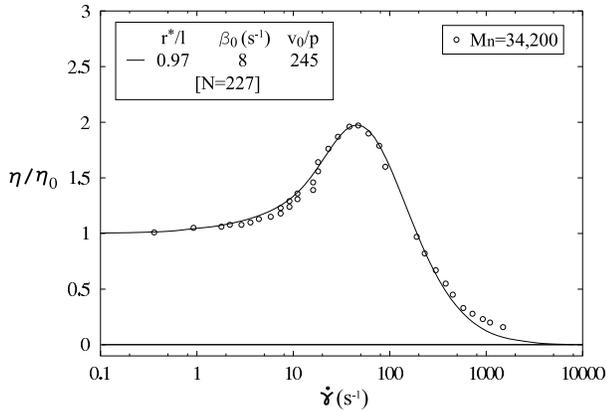}
\end{center}
\caption{
The steady shear viscosity 
plotted against the shear rate for $N\!=\!227$ ($M_n\!=\!34,200$).
The solid line represents the theoretical result, whereas
the dots represent the experimental results observed for
$\eta_0\!=\!3.0$Poise, 
$T\!=\!298$K and $c\!=\!1$wt\%.~\protect\cite{Jenkins1}
} 
\label{Nloop22pv}
\end{figure}

\begin{figure} 
\begin{center}
\includegraphics*[scale=0.6]{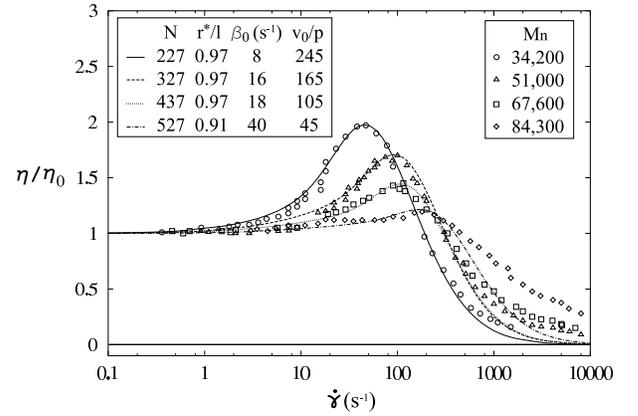}
\end{center}
\caption{The steady shear viscosity plotted against the shear rate.
The lines represent theoretical results obtained for
each $M_n$ listed in Table~\protect\ref{hyoup2}.
The dots represent the experimental results observed for 
$T\!=\!298$K and $c\!=\!1$wt\%.~\protect\cite{Jenkins1} 
The zero-shear viscosity $\eta_0$ is 
$3.0$Poise for $M_n\!=\!34,200$,
$0.94$Poise for $M_n\!=\!51,000$,
$0.53$Poise for $M_n\!=\!67,600$ and
$0.26$Poise for $M_n\!=\!84,300$.~\protect\cite{Jenkins1}} 
\label{Nloopv}
\end{figure}

The steady shear viscosity for $r^*\!=\!0.97l$, $\beta_0\!=\!8$s$^{-1}$ 
and $v_0/p\!=\!245$ ($N\!=\!227$) 
is shown in Figure \ref{Nloop22pv}.
We see that the calculated viscosity coincides well with
experimental data for the broad shear rate.
Note that the value $r^*\!=\!0.97l$ is quite close to
the value $0.96l$ derived from the consideration based on the potential barrier
between a junction and a hydrocarbon end group. 
In the figure, we see   
the gradual enhancement of the viscosity at lower shear rates.
This is because 
the number of active chains increases with shear rate in this region 
as shown in Figure \ref{Nloop22nupv0}.
In addition, there is the effect of stretched active chains.
It adds the sharp enhancement in the viscosity at the shear rate where
the end-to-end length of the active chain reaches $r^*$.
When loops are absent, 
we had to adopt $r^*$ very close to $l$ 
(see appendix \ref{appeTElimit}).\cite{Master}
However, in the present case, since the increase in the number of
active chains at lower shear rates promotes the 
viscosity rising, we can adopt proper value for $r^*$.

One might suppose that the observed shear viscosity can be fit 
without considering the high extension of active chains.
However, it is impossible.
When $r^*\!=\!0.5l$, for instance, we cannot heighten the 
peak, up to the height of observed one with keeping the 
relation (\protect\ref{kansanp}) or (\ref{kansanpv0}).
Therefore, in the system we are dealing with,
the above two mechanisms 
(i.e., the enhancement of the number of active chains and 
the high-extension of active chains) are indispensable.

The steady shear viscosity are calculated as shown in Figure \ref{Nloopv}
with optimal values of $r^*$, $\beta_0$ and $v_0/p$ 
(given inside the box in the figure) 
for each molecular weight listed in Table \ref{hyoup2}.
We see from the figure that the theoretical curve for each 
molecular weight $M_n$ fits very well with the first peak  
(although when $M_n\!=\!84,300$, it does not fit well at higher shear rate). 
Note that $r^*$ does not depend on $N$, except for  
$N\!=\!527$.

For each $v_0/p$, the values of 
$p$ and $v_0$ are given via (\ref{kansanpv0}), e.g.,
$p\!=\!0.14$s$^{-1}$, $v_0\!=\!35$s$^{-1}$ 
for $v_0/p\!=\!245$ ($N\!=\!227$),
$p\!=\!0.1$s$^{-1}$, $v_0\!=\!17$s$^{-1}$ for $v_0/p\!=\!165$ ($N\!=\!327$)
and $p\!=\!0.065$s$^{-1}$, $v_0\!=\!6.8$s$^{-1}$ 
for $v_0/p\!=\!105$ ($N\!=\!437$).
Assuming that $p$ and $v_0$ obey, respectively, the scaling low
with respect to $N$, i.e., $p\propto N^{-\epsilon}$ and 
$v_0\propto N^{-\zeta}$,
we see from these data that $\epsilon\!\simeq\!1.2$ and $\zeta\!\simeq\!2.5$
by the method of least squares.
The dependence of $p$ on $N$ may be interpreted as follows. 
Since the polymer concentration is fixed,
the number of hydrophobic end groups within the system decreases 
as the molecular weight increases. 
It diminishes the chance for a dangling chain to become an active chain.
For example, when the molecular weight is doubled,
the number of end groups within the system is reduced by half. 
This suggests $p\!\propto \!N^{-1}$ 
giving $\epsilon\!=\!1$, hence 
the value of the exponent $\epsilon\!\simeq\!1.2$, 
obtained by the present analysis, is fairly reasonable.
The dependence of $v_0$ on $N$ is also fairly reasonable
since the probability for a self-avoiding chain to form a loop is 
proportional to $N^{-2}$ giving $\zeta\!=\!2$.\cite{degenneshon} 
Recall, however, that we used just three data to determine the exponents.
Obviously, we need more data 
in order to discuss further details of the $N$ dependence of $p$ and $v_0$.
It is desirable that some experiments will be conducted 
to provide enough number of data for 
different molecular weights which may 
support a more precise analysis on the $N$ dependence.


\section{Summary and Discussion} 
\label{discoz}

We developed the network model for physical gel composed of
unentangled linear chains with associative functional groups at 
both their ends.
The presence of three types of chains was assumed in the network system:
active chains with both their ends attached to the network responsible for
the stress of the system, dangling chains with one of their two ends 
sticking to
the network and loops with both their ends connecting to a junction. 
We investigated the transition rates between these chains
how they depend on the end-to-end vector $\br$
and on the shear rate $\dg$.
The introduction of cutoff length $r^*$ was shown to be essential 
for the treatment of stretched active chains. It was also revealed that 
the transition from a loop 
to a dangling chain is promoted by
the shear flow because
of the collision with 
dangling chains or other loops.

On the basis of the above network model,
we showed that the origin of shear-thickening behavior is
ascribed to 1) the dissociation of loops enhanced by shear flow, 
and 2) the high extension of active chains promoted by the flow.
The portion of low and broad peak with weak $N$-dependence 
originates from the mechanism 1), and the portion of
high and narrow peak with strong $N$-dependence 
stems from the mechanism 2).

\begin{figure}
\begin{center}
\includegraphics*[scale=0.6]{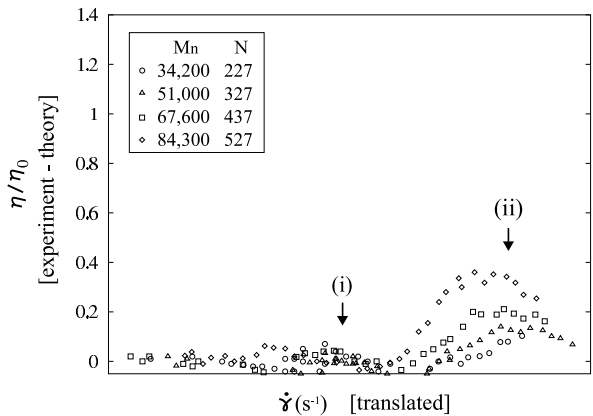}
\end{center}
\caption{The difference between experimentally and theoretically
obtained shear viscosity.
The theoretical results 
(Figure \protect\ref{Nloopv})
are subtracted from experimental data. 
To clarify the $N$ dependence of the height of peaks,
each data is translated so that the positions of the observed peaks 
coincide with each other.
See also Figure \protect\ref{TEgap} for a reference.}
\label{waregap}
\end{figure}

It is revealed that the first peak in Figure \ref{TEgap} can 
be well explained by these two mechanisms.
In Figure \ref{waregap}, we put the difference between
theoretically obtained shear viscosity 
(Figure \ref{Nloopv})
and the observed shear viscosity. 
The disappearance of the first peak in the difference declares that
we succeeded to explain the shear-thickening phenomenon
by the present theoretical analysis.

In our analysis, we took the limit $\tau\!\to\!0$ for the
relaxation time of dangling chains, since the characteristic time
of the shear flow is around $10^{-3}\!\sim\!10^{-1}$s, while
$\tau$ is estimated as $10^{-5}\!\sim\!10^{-4}$s (see Table \ref{hyoup3}).
Within the limit $\tau\!\to\!0$, 
an active chain which detaches its one end
from the network relaxes 
to the Gaussian chains\footnote{
At equilibrium state, the conformation of a dangling chain is 
regarded as the Gaussian.}
in an instant, and never rejoins to the network during the relaxation.
If the finite relaxation time 
are taken into account, on the contrary, the detached chain
has the possibility to recombine with the network before
it totally relaxes to the Gaussian chain.\cite{Mar1,Mar0} 
Note that $\tau$ is larger for larger $N$ (see (\ref{deftau})),
implying that the detached chain with larger $N$ 
has higher possibility to recombine with the other
junction during its relaxation.
It indicates that the chains with larger molecular weight
contribute more to the viscosity around the shear rate
where characteristic time of the flow is the same extent
as that of dangling chains.
We suppose that this mechanism may explain the second peak
(denoted by (ii) in Figure \ref{TEgap} and Figure \ref{waregap}),
since it appears at higher shear rates ($\dg\gnear10^3$s$^{-1}$), 
and the larger the molecular weight is, 
the taller the height of the peaks becomes. 
To include these mechanisms into the present formula 
is one of the interesting future problems
and will be reported elsewhere.

Finally, let us comment on the idea of `shear-induced transfer
of intramolecular 
to intermolecular associations',
proposed and investigated by Witten and Cohen~\cite{WC} 
in qualitative ways, and developed by Ballard {\it et al.}
in quantitative manners.\footnote{
Their discussions are not based on the transient network theory.}
They considered solutions of associative polymers having 
some functional groups along chains.
When a flow is not added to the solution, 
each chain is in the random coil state, and hence, 
has a number of associations among functional 
groups {\it inside} the chain (intramolecular associations).
On the other hand, when sufficiently large velocity gradient
is added to the solution, the chain extends due to 
viscous interactions with the solvent, 
thereby breaking there intramolecular associations.
This causes associations among functional groups belonging to
{\it different} chains (intermolecular associations), 
and enhances the viscosity.
If the number of functional groups is two, and these 
functional groups are attached at each end of a chain, 
the intramolecular association makes a loop.
In this case, shear-thickening is caused by
`shear-induced transfer of loops to active chains via dangling chains' 
in their interpretation.

One might say that this process is the same as the process proposed 
in the present paper. 
However, the origin to open a loop is quite different, i.e.,
as shown in section \ref{characchap}, 
the dissociation of a loop is caused by 
collisions with other chains, in the present model.
Indeed, viscous interactions with solvent
affect the conformation of loops (see (\ref{loopnopx})),
and enhance the possibility for a loop to be a dangling chain, however, 
this works effectively only at 
higher shear rates ($\dg\!\gnear\!1/\tau_1\!\sim\!10^{4}$s$^{-1}$).
Therefore, it seems that 
the dissociation of loops caused by viscous interactions
do not contribute to the shear-thickening behavior 
at the moderate shear rates,
although they may influence the second peak in Figure \ref{waregap}.
To include the mechanism\footnote{
That is, the shear-induced transfer of 
loops to active chains via dangling chains,
by means of {\it frictional force due to solvent}.}
into our model is also interesting future problems.


\appendix

\section{Radius of gyration of a loop}
\label{appenradi}

We use the sine transform for $\bx(n,t)$ 
so as to satisfy the boundary condition (\ref{bankon0}), i.e.,
\bea
\bx(n,t)=2\sum_{p=1}^{\infty}\bx_p(t)\sin \Biggl(\frac{n\pi}{N}p\Biggr).
\label{fourrrr}
\eea
We also use the sine transform for $\bR(n,t)$, i.e.,
\bea
\bR(n,t)=2\sum_{p=1}^{\infty}\bR_p(t)\sin \Biggl(\frac{n\pi}{N}p\Biggr).
\label{fourer}
\eea
Due to (\ref{rloopj1}) and (\ref{rloopj2}), 
the average and the variance of $\bR_p(t)$ are given by
\bea
& &\bra\!\bra R_{p\alpha}(t)\ket\!\ket=0 ,\label{rloopj11}\\
& &\bra\!\bra R_{p\alpha}(t)R_{q\beta}(s)\ket\!\ket=
\frac{\zeta_1k_BT}{N}\delta_{\alpha \beta}\delta_{pq}
\delta(t-s),\label{rloopj22}
\eea
respectively.
Substituting (\ref{fourrrr}) and (\ref{fourer}) into 
(\ref{lanlosaren}), we obtain equations for $\bx_p(t)$ as follows:
\bea
\frac{d\bx_p(t)}{dt}=\Biggl(\hat{\kappa}(t)-\frac{1}{\tau_p}\Biggr)\bx_p(t)
+\frac{1}{\zeta_1}\bR_p(t), 
\label{loopbiho}
\eea
where we put relaxation time of `mode' $p$ as 
\bea
\tau_p=\frac{\zeta_1 N^2}{K_1\pi^2}\frac{1}{p^2}
=\frac{\zeta_1N^2a^2}{3\pi^2k_BT}\frac{1}{p^2} \label{raustaup}.
\eea
Integrating (\ref{loopbiho}), from time $-\infty$ to $t$, 
we have
\bea
\bx_{p}(t)=\frac{1}{\zeta_1}\int_{-\infty}^{t}\hspace*{-0.2cm}dt^{\prime}~
\e^{-(t-t^{\prime})/\tau_p}\hat{\lambda}(t,t^{\prime})\bR_{p}(t^{\prime}).
\hspace*{-0.3cm}\label{loopoketa}
\eea
Making use of (\ref{rloopj22}), the correlation functions of 
$x_{p\alpha}(t)$ are written as
\bea
& &\bra\!\bra x_{p\alpha}(t_1)x_{q\beta}(t_2)\ket\!\ket 
=\delta_{pq}\frac{k_BT}{\zeta_1N}
\int_{-\infty}^{\min{(t_1,t_2)}}\hspace*{-0.5cm}dt^{\prime}
\e^{-(t_1+t_2-2t^{\prime})/\tau_p} \non\\
& &\hspace*{3cm}\times \sum_{\gamma=1}^3\lambda_{\alpha\gamma}(t_1,t^{\prime})
\lambda_{\beta\gamma}(t_2,t^{\prime}).\label{loopoketa22}
\eea
For the steady shear flow represented by the deformation tensor 
(\ref{shearlam}), we have
\bea
\hspace*{-0.8cm}\dbra x_{px}(t_1)x_{qx}(t_2)\dket
&=&\delta_{pq}\frac{Na^2}{6\pi^2}\frac{1}{p^2}\e^{-|t_1-t_2|/\tau_p} \non\\
& &\hspace*{-1.cm}\times
\Biggl( 1+\frac{1}{2}\tau_p\dg^2|t-s|+\frac{1}{2}(\tau_p\dg)^2\Biggr), 
\label{radisx}  \\
\dbra x_{py}(t_1)x_{qy}(t_2)\dket&=&\dbra x_{pz}(t_1)x_{qz}(t_2)\dket \non\\
&=&\delta_{pq}
\frac{Na^2}{6\pi^2}\frac{1}{p^2}\e^{-|t_1-t_2|/\tau_p}, \label{radisy}\\
\dbra x_{px}(t_1)x_{qy}(t_2)\dket&=&\delta_{pq}\frac{Na^2}{6\pi^2}
\frac{1}{p^2}\e^{-|t_1-t_2|/\tau_p} \non\\
& &\hspace*{-1cm}
\times\dg\Biggl( (t_1-t_2)\theta(t_1-t_2)+\frac{\tau_p}{2}\Biggr), \\
\dbra x_{px}(t_1)x_{qz}(t_2)\dket&=&\dbra x_{py}(t_1)x_{qz}(t_2)\dket=0.
\label{radisxee}
\eea
Putting
\bea
\bx_G(t)\!=\!\frac{1}{N}\!\int_0^N \!\!\!dn~\bx(n,t) 
\!=\!\frac{4}{\pi}\sum_{p:\mbox{{\scriptsize odd integer}}}^{\infty}
\!\frac{\bx_p(t)}{p} \label{cengra}
\eea
into (\ref{loopnpse}), we get
\bea
s_{\alpha}s_{\beta}
&=&2\sum_{p=1}^{\infty} \dbra x_{p\alpha}(t)x_{p\beta}(t)\dket \non\\
& &-\frac{16}{\pi^2}
\sum_{p,q:\mbox{{\scriptsize odd integer}}}^{\infty}
\frac{\dbra x_{p\alpha}(t)x_{q\beta}(t)\dket}{pq}.\label{radigs}
\eea
Substituting (\ref{radisx}) $\sim$ (\ref{radisxee}) 
into (\ref{radigs}), we obtain
(\ref{loopnopx}) $\sim$ (\ref{ynoradsonota}).


\section{Derivation of the Fokker-Planck Equation} 
\label{sttt2}

Since $\phi_f^i(\br,t)$ is a stochastic variable, its
differentiation with respect to time is not well-defined.
So, we should replace the equations 
(\ref{loeusyo00}) $\sim$ (\ref{loeusyo})
by the forms of stochastic differential equations.
As for (\ref{dsli}), we have
\bea
d\phi_f^d(\br,t)=d\Omega(\br,t)\circ\phi_f^d(\br,t)+\Psi_f(\br,t)dt,
\label{stchasone}
\eea
where $\Omega(\br,t)dt$ is the {\it stochastic operator} defined by
\bea
& &\hspace*{-0.4cm}d\Omega(\br,t)=-\nabla\!\cdot\! \Biggl[\Biggl(
\hat{\kappa}(t)\!-\!\frac{K_N}{\zeta_N}\Biggl)\br dt+\frac{1}{\zeta_N}
d\bw(t)\Biggr], \label{omgagas}
\eea
and 
\bea
& &\hspace*{-0.5cm}
\Psi_f(\br,t)=\beta_0\phi_f^a(\br,t)\!-\!(p\!+\!v(\br))\phi_f^d(\br,t)\!+\!
u\phi_f^l(\br,t).
\eea
$\bw(t)$ in (\ref{omgagas}) is called the {\it Wiener process}. 
The correlations of its increment $d\bw(t)$ are given by
\bea
& &\bra\!\bra dw_{\alpha}(t)\ket\!\ket=0, \label{wiener0}\\
& &\bra\!\bra dw_{\alpha}(t)dw_{\beta}(s)\ket\!\ket=2\zeta_Nk_BT
\delta_{\alpha\beta}\delta (t-s)dtds. \label{moment2wi}
\eea
Note that the Wiener process $\bw(t)$ is related with 
the Gaussian white process $\bR(t)$ through
\bea
\bw(t)=\int_0^tdt^{\prime}~\bR(t^{\prime}).
\eea
Using the fact
\bea
dw_{\alpha}(t)dw_{\beta}(t)\!=\!2\zeta_Nk_BT\delta_{\alpha\beta}dt,
\label{proconve}
\eea
we obtain 
\bea
d\Omega(\br,t)d\Omega(\br,t)
&=&\sum_{\alpha,\beta}\nabla_{\alpha}\nabla_{\beta}
\frac{1}{\zeta_N^2}dw_{\alpha}(t)dw_{\beta}(t)\non\\
&=&\nabla^2\frac{2k_BT}{\zeta_N}dt,\label{isotoroo2}
\eea 
where we neglected the terms of order higher than $dt$ 
for the first equality.
Note that the product rule (\ref{proconve}) holds in the sense of 
{\it stochastic convergence}.\cite{namiki}
The relation (\ref{isotoroo2}) is the content of
the {\it fluctuation-dissipation theorem of the second kind}.
Hereafter, we will omit the argument $\br$ 
in $\phi_f^d(\br,t)$, $\Psi_f(\br,t)$ and $\Omega(\br,t)$ 
for simplicity.

In (\ref{stchasone}), there appears the product of
two stochastic variables, i.e., $\phi_f^d(t)$ and $d\bw(t)$.
There are two types of such product: 
one is the product of the {\it Ito} type\cite{itob} which is defined by
\bea
\phi_f^d(t)\cdot d\bw(t)=
\phi_f^d(t)[\bw(t\!+\!dt)\!-\!\bw(t)], \label{itose}
\eea
and the other is that of the {\it Stratonovich} type\cite{strat} 
defined by
\bea
\hspace*{-0.7cm}
\phi_f^d(t)\circ d\bw(t)\!=\!\frac{\phi_f^d(t\!+\!dt)\!+\!\phi_f^d(t)}{2}
[\bw(t\!+\!dt)\!-\!\bw(t)].\label{sttose}
\eea
Since $\phi_f^d(t)$ is independent of 
the future stochastic processes $\bw(t\!+\!dt)\!-\!\bw(t)$, 
the Ito product has a notable property
\bea
\bra\!\bra\phi_f^d(t)\cdot d\bw(t)\ket\!\ket=
\bra\!\bra\phi_f^d(t)\ket\!\ket\bra\!\bra d\bw(t)\ket\!\ket=0,
\label{itoprop}
\eea
where we used (\ref{wiener0}) for the second equality. This property often
simplifies our calculations.
From the definitions (\ref{itose}) and (\ref{sttose}),
we have the formula connecting the Stratonovich product
with the Ito product in the form
\bea
\hspace*{-0.2cm}
\phi_f^d(t)\circ d\bw(t)=\phi_f^d(t)\cdot d\bw(t)\!+\!\frac{1}{2}
d\phi_f^d(t) d\bw(t). \label{formula}
\eea
As stated in subsection \ref{stprosec}, 
we regard the stochastic product appeared in (\ref{stchasone})
as that of the Stratonovich type.

Applying the formula (\ref{formula}) to 
$d\Omega(t)\circ\phi_f^d(t)$ in (\ref{stchasone}), and 
neglecting the terms of order higher than $dt$, 
we have the stochastic equation of the Ito type  
\bea
\hspace*{-0.7cm}d\phi_f^d(t)\!
&=&\!d{\it \Omega}(t)\cdot\phi_f^d(t)\!+\!\Psi_f(t)dt, \label{itosthen}
\eea
where
\bea
d{\it \Omega}(t)&=&d\Omega(t)+\frac{1}{2}d\Omega(t)d\Omega(t) \non\\
&=&d\Omega(t)+\nabla^2\frac{k_BT}{\zeta_N}dt  \non\\ 
&=&-\nabla\!\cdot\! \Biggl[\Biggl(
\hat{\kappa}(t)\!-\!\frac{K_N}{\zeta_N}\Biggl)\br 
\!-\!\frac{k_BT}{\zeta_N}\nabla\Biggr]dt\!-\!\frac{1}{\zeta_N}
\nabla\!\cdot\!d\bw(t). \label{stprosec2} \non\\
\eea 
Taking the random average of the stochastic equation (\ref{itosthen}),
we obtain the Fokker-Planck equation
\bea
d\phi^d(t)&=&-\nabla\!\cdot\! \Biggl(\Biggl[\Biggl(
\hat{\kappa}(t)\!-\!\frac{K_N}{\zeta_N}\Biggl)\br\!-\!\frac{k_BT}{\zeta_N}
\nabla\Biggr] \phi^d(t) \Biggr)dt \non\\
& &+ \Psi(t)dt, \label{saigoobt}
\eea
where $\phi^i(t)\!=\!\bra\!\bra \phi^i_f(t)\ket\!\ket$ 
as defined in subsection \ref{stprosec}.
In order to obtain (\ref{saigoobt}) from (\ref{itosthen}),
we used the property of the Ito product (\ref{itoprop}).

As for the equations (\ref{loeusyo00}) and (\ref{loeusyo}),
since they do not contain the stochastic process $\bR(t)$ explicitly,
we can obtain (\ref{loeusyof01}) and (\ref{ljfk}) in a straightforward way
by taking the random average of (\ref{loeusyo00}) and (\ref{loeusyo}),
respectively.


\section{Shear Viscosity (without Loops)} 
\label{appeTElimit}

\subsection{TE Limit}

\begin{figure} 
\begin{center}
\includegraphics*[scale=0.6]{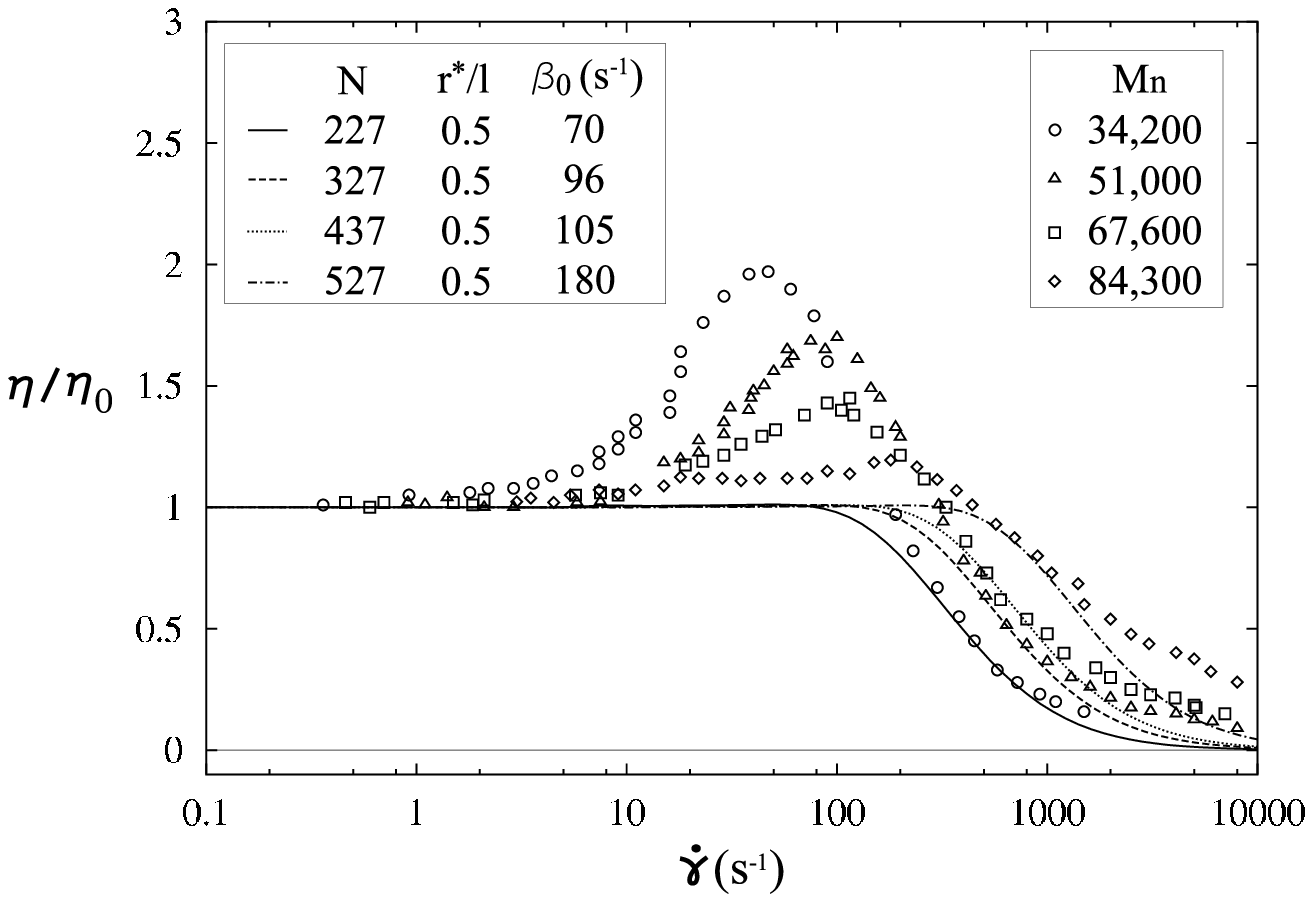}
\end{center}
\caption{
The steady shear viscosity plotted against the shear rate (TE limit).
The lines represent theoretical (TE limit) results obtained for
each $M_n$ listed in Table~\protect\ref{hyoup2} 
with $r^*/l\!=\!0.5$.
The dots represent the experimental results observed for 
$T\!=\!298$K and $c\!=\!1$wt\%.~\protect\cite{Jenkins1} 
The zero-shear viscosity $\eta_0$ is 
$3.0$Poise for $M_n\!=\!34,200$,
$0.94$Poise for $M_n\!=\!51,000$,
$0.53$Poise for $M_n\!=\!67,600$ and
$0.26$Poise for $M_n\!=\!84,300$.~\protect\cite{Jenkins1}}
\label{TEviss}
\end{figure}

Here, we compare the steady shear viscosity observed by
Jenkins {\it et al.}~\cite{Jenkins1} with that given by the TE limit
(Figure \ref{TEviss}) for each molecular weight 
listed in Table \ref{hyoup2}.
As mentioned in subsection \ref{secTE},
the TE limit is attained by adopting small $r^*$ and putting $v_0\!=\!0$
in equation (\ref{saigonouryo}).
Since $p$ is given through (\ref{kansanp}) or (\ref{kansanpv0}),
the number of adjustable parameters is two, i.e., $\beta_0$ and $r^*$.

\begin{figure}[ht]
\begin{center}
\includegraphics*[scale=0.6]{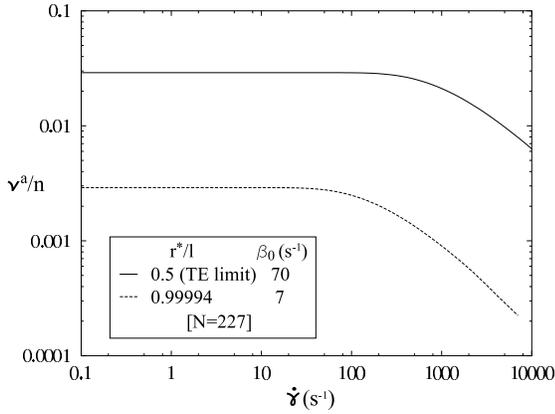}
\end{center}
\caption{The number of active chains 
plotted against the shear rate for $N\!=\!227$. 
The solid line stands for the number of active chains 
within the TE limit which corresponds to
the solid line in Figure \protect\ref{TEviss}. 
The dashed line represents the number of active chains 
corresponding to the solid line 
in Figure \protect\ref{noNloopv}.} 
\label{Nloop22nupv}
\end{figure}

We adopt $r^*\!=\!0.5l$ here,
although the value is not so small enough being good for the Gaussian model
or the Hookean force assumption.\footnote{
This reason is as follows. For smaller $r^*$, 
the viscosity starts to decrease at lower shear rates, 
since the lifetime of an active chain is shorter effectively in this case.
It means that, for smaller $r^*$, 
we have to adopt larger $\beta_0$ (smaller $1/\beta_0$) 
so that the theoretical results fit to experimental data
at the shear-thinning region.
Therefore, in order to get larger $1/\beta_0$ close to the observed 
relaxation time, 
we have to choose larger $r^*$.
Although the value $r^*\!=\!0.5l$ generates 
stronger force
compared with Hookean force, it does not cause viscosity rising. 
This value seems to be the critical one 
for generating the
shear-thickening behavior (see Figure \ref{noNloopN}).} 
With this $r^*$, the value of $\beta_0$ is determined by
adjusting the theoretical curve to experimental data
{\it at the shear-thinning region}
(the abscissa is scaled by $\beta_0$ for given $N$).
For example, 
the adjusted value of $\beta_0$ is $70$s$^{-1}$
for $N\!=\!227$ ($M_n=34,200$)
giving us
the lifetime $1/\beta_0\!\simeq\!0.01$s of active chains.
It is much smaller than the observed relaxation time of HEUR
aqueous solutions having the order of 
0.1s.~\cite{Jenkins1,Annable1}
The values of $\beta_0$ for other $N$ are given inside the box 
in Figure \ref{TEviss}.

We see from these figures that 
the shear viscosity decreases monotonically as the shear rate increases,
i.e., the shear-thickening behavior 
does not appear within the TE limit.
With increasing the shear rate, the number of {\it longer} 
active chains increases.
The chains which exceed $r^*$ dissociate from the junction and become dangling 
chains.
This causes the decrease in the number of active chains 
(see the solid line in Figure \ref{Nloop22nupv}.)
leading to the thinning in the shear viscosity.
Since $r^*$ is not close to $l$ within the TE limit, 
the tension of the active chain
with end-to-end length $r^*$ is not so strong.
Therefore, they can not generate the enhancement of the viscosity.
That is the reason why the viscosity rise does not occur in the TE limit.


\subsection{Effects of High-Extension of Active Chains}\label{efhiten}

\begin{figure}
\begin{center}
\includegraphics*[scale=0.6]{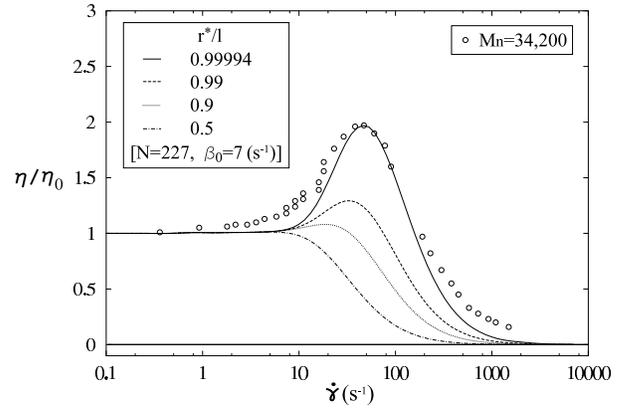}
\end{center}
\caption{
The steady shear viscosity plotted against the shear rate (without loops). 
The lines represent theoretical results ($v_0\!=\!0$) obtained for
$r^*/l\!=\!0.5$, $0.9$, $0.99$ and $0.99994$ with 
fixed $N\!=\!227$ and $\protect\beta_0\!=\!7 $s$^{-1}$. 
The dots represent the experimental results observed for 
$M_n\!=\!34,200$,
$\eta_0\!=\!3.0$Poise, 
$T\!=\!298$K and $c\!=\!1$wt\%.~\protect\cite{Jenkins1}}
\label{noNloopN}
\end{figure}

\begin{figure}
\begin{center}
\includegraphics*[scale=0.6]{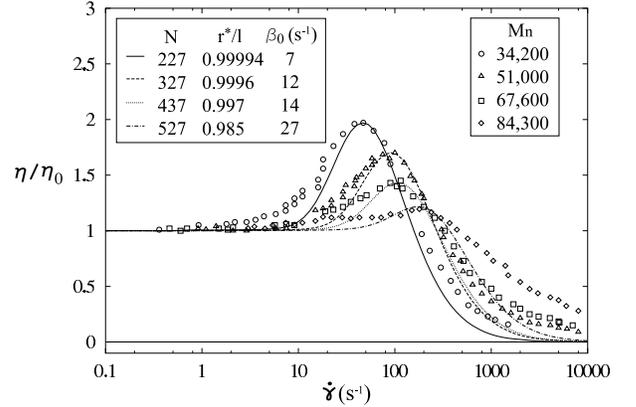}
\end{center}
\caption{The steady shear viscosity plotted against the shear rate 
(without loops.)
The lines represent theoretical results ($v_0\!=\!0$) obtained for
each $M_n$ listed in Table~\protect\ref{hyoup2}. 
The dots represent the experimental results observed for 
$T\!=\!298$K and $c\!=\!1$wt\%.~\protect\cite{Jenkins1} 
The zero-shear viscosity $\eta_0$ is 
$3.0$ for $M_n\!=\!34,200$,
$0.94$ for $M_n\!=\!51,000$,
$0.53$ for $M_n\!=\!67,600$ and
$0.26$ for $M_n\!=\!84,300$.~\protect\cite{Jenkins1}} 
\label{noNloopv}
\end{figure}

Let us now see the influence of the stretched active chains on the viscosity.
To this end, we consider the case where loops are absent ($v_0\!=\!0$)
in the wake of the preceding subsection.

The $r^*$ dependence of the steady shear viscosity  
is shown in Figure \ref{noNloopN} for $N\!=\!227$ ($M_n\!=\!34,200$) and
$\beta_0\!=\!7 $s$^{-1}$.
From Figure \ref{noNloopN}, we see that a peak
starts to appear when $r^*\!\gnear\!0.5l$,
and the height of the peak becomes higher with increasing $r^*$.
When $r^*\!=\!0.99994l$, 
it coincides with the experimentally observed one. 
Note that the appeared peak is narrower than observed one.
The value $\beta_0\!=\!7 $s$^{-1}$ is determined 
so that the position of the peak coincides with 
observed one.
The obtained $1/\beta_0\!=\!0.14$s$^{-1}$ 
has the same order as 
observed relaxation time 
of HEUR aqueous solutions.~\cite{Jenkins1,Annable1}

Figure \ref{noNloopv}
shows the theoretically obtained steady shear viscosity 
with optimal values of $r^*$ and $\beta_0$ (given inside the box in the figure)
for each molecular weight 
listed in Table \ref{hyoup2}, compared with experiments.
We see from these figures, in common, 
that 1) the cutoff length $r^*$ is too close to the contour length $l$,
and 2) the peak is too sharp.

The number of active chains for $N\!=\!227$ is shown in 
Figure \ref{Nloop22nupv} (the dashed line).
As one can see, it 
monotonically decreases with increasing the shear rate, 
the reason of which  
is the same as in the case of $r^*\!=\!0.5l$ (i.e., TE limit).
In order to enhance the viscosity against the decreasing
number of active chains, we are forced to choose $r^*$ very close to $l$,
since $r^*\!\lnear\!l$ causes the strong force between ends of 
active chains.
That is why $r^*$ so near to $l$ is required to fit the experimental data.
The above problems 1) and 2) is resolved by taking account of 
the open process of loops.


\end{document}